\documentclass[conference]{IEEEtran}
\ifCLASSINFOpdf
\else
\fi
\usepackage{graphicx}

\usepackage{amsmath}
\usepackage{caption}
\usepackage{subcaption}
\usepackage{url}
\usepackage[linesnumbered,ruled,vlined]{algorithm2e}

\begin{document}

%
\title{\textbf{Low-Latency High-Level Data Sharing for Connected and Autonomous Vehicular Networks}}

\author{
  \IEEEauthorblockN{
    Qi Chen\IEEEauthorrefmark{1},
    Sihai Tang\IEEEauthorrefmark{1},
    Jacob Hochstetler\IEEEauthorrefmark{1},
    Jingda Guo\IEEEauthorrefmark{1},
    Yuan Li\IEEEauthorrefmark{1},
    Jinbo Xiong\IEEEauthorrefmark{2},
    Qing Yang\IEEEauthorrefmark{1},
    Song Fu\IEEEauthorrefmark{1}
  }
  \IEEEauthorblockA{
    \IEEEauthorrefmark{1}\textit{Department of Computer Science and Engineering,}
    \textit{University of North Texas, USA}\\
    \IEEEauthorrefmark{2}\textit{College of Mathematics and Informatics,}
    \textit{Fujian Normal University, China}\\
  }
  \IEEEauthorblockA{
    \textit{\{QiChen, SihaiTang, JacobHochstetler, JingdaGuo, YuanLi4\}@my.unt.edu}
    \textit{\{Jinbo.Xiong, Qing.Yang, Song.Fu\}@unt.edu}
  }
}


%


\maketitle
\begin{abstract}
Autonomous vehicles can combine their own data with that of other vehicles to enhance their perceptive ability, and thus improve detection accuracy and driving safety. 
Data sharing among autonomous vehicles, however, is a challenging problem due to the sheer volume of data generated by various types of sensors on the vehicles.
In this paper, we propose a low-latency, high-level (L3) data sharing protocol for connected and autonomous vehicular networks.
Based on the L3 protocol, sensing results generated by individual vehicles will be broadcasted simultaneously within a limited sensing zone.
The L3 protocol reduces the networking latency by taking advantage of the capture effect of the wireless transmissions occurred among vehicles. 
With the proposed design principles, we implement and test the L3 protocol in a simulated environment.
Simulation results demonstrate that the L3 protocol is able to achieve reliable and fast data sharing among autonomous vehicles.

\end{abstract}


\section{Introduction}
For years, the development of connected and autonomous vehicles (CAV) technology has garnered significant interest from both research institutes and industry alike.
CAV incorporate a variety of different technologies, ranging from computer vision~\cite{geiger2012we} to wireless networking~\cite{Yang-2010}, to facilitate a safe and efficient movement of people and goods, revolutionizing the current transportation system.
It brings a host of benefits such as improved safety, convenient mobility for the elderly and disabled, and a better public transportation system~\cite{icme2019}. 
Ideally, CAV could help drive fatalities to near zero, given the technologies continue to improve.

Autonomous vehicles are typically equipped with high-precision sensing systems, producing a healthy amount of sensor data that need to be processed in real time.
For example, the autonomous vehicles developed by companies such as Google, Tesla, Mobileye autopilot systems, are mainly equipped with LiDAR (light detection and ranging), camera, Radar, ultrasound, thermal camera, GPS, and IMU (inertial measurement unit ), etc.
Currently, the data generated by these sensors are processed and stored locally on individual vehicles, and rarely shared among autonomous vehicles.
The current solutions, however, come with several limitations. 
When driving in the evening, rain, snow, fog or other bad weather conditions, cameras may not work properly.
Similarly with LiDAR and Radar sensors, intersections, turning corners and other scenarios may witness the sensing systems not functioning properly.
For example, Tesla autonomous car once had a fatal accident on a freeway.
The vehicle failed to identify the white body of a truck under an intense sunshine condition, and therefore did not activate the brake system in time.
While developing more powerful sensors may solve these issues, the associated cost will rise to the point where consumers cannot afford the product, i.e., individual customers aren't likely to see such vehicles in much volume.

A possible solution to the above mentioned issues is to allow autonomous vehicles to exchange real-time sensing data to each other, realizing a connected and autonomous vehicular system.
Although data sharing among vehicles is promising, it faces several challenges that must be adequately addressed before the technology is deployed in real world.
These challenges are related to both data processing and data sharing, e.g., it is difficult to synchronize the sensing data among vehicles, the networking bandwidth of existing wireless technologies is too limited to transmit the data, the large networking delay may be prohibitive for autonomous driving applications.
In summary, without efficient data processing and transmission mechanisms, the sheer amount of resources that will be consumed by autonomous vehicles can dramatically slow the deployment of CAV technologies.

\subsection{Motivations}
To achieve data sharing among autonomous vehicles, an effective way is to adopt the V2X communications~\cite{yang2016architecture}.
Intuitively, V2X communication connects cars to other cars or the Internet to form a vehicular networks, including V2V (vehicle to vehicle), V2I (vehicle to infrastructure), V2N (vehicle to Internet) and V2P (vehicle to people) communications.
V2X communication can be viewed as a means that allows the sensors on vehicles to extend their sensing range well beyond what they are physically capable of.
By sharing the sensing results with nearby vehicles and roadside infrastructures, vehicles can greatly enhance the perception of the surrounding environment and thus enhancing their decision making.
Even though the self-driving function can be partially achieved by the vehicle itself, using V2X can further improve safety and driving performance by reducing the cost of deploying high-precision sensors.
In addition to improving its own perception and decision making, the enabled autonomous vehicle can also improve the driving reliability for the normal human operated vehicles, making it more encouraging for the adoption of more vehicles equipping V2X devices.

\subsection{Challenges}
The challenge of a CAV system comes from the massive deployment of sensors on the autonomous vehicles and the huge amount of data that they can pick up from the environment~\cite{yick2008wireless}.
First of all, it is challenging to synchronize and fuse data generated from different vehicles which may use different sensors (and algorithms) to perceive the surrounding environment.
Data fusion is the process of combining multiple vehicles' data to produce a more consistent, accurate, and reliable perception than what is offered by an individual vehicle~\cite{fusion}.
The data fusion process in CAV is usually classified into three categories: low-level, intermediate-level, and high-level fusions, depending on the processing stage at which the fusion takes place~\cite{level}.
As their names imply, low-level fusion refers to raw data fusion, which requires the highest network bandwidth to transmit the data.
Intermediate-level fusion, such as feature-based fusion, takes the features extracted from the raw data before fusion. 
Finally, high-level fusion takes the processing results, e.g., the objects detected from cameras, to conduct the fusion.
For wireless vehicular networks, regardless which type of fusion is adopted, the large amount of data generated and shared among vehicles will pose significant research challenges to existing wireless technologies, e.g., dedicated shorte range communication (DSRC)~\cite{dsrc} and 5G networks.

\subsection{Proposed Solution}
To facilitate data sharing among autonomous vehicles, high-level fusion is often opted over the other two levels of fusion, due to the small amount of data exchanged between vehicles. 
In this way, each vehicle processes its sensing data locally and exchange its sensing results with nearby vehicles.
As long as the format of the sensing results are standardized, it does not matter what sensing technologies individual vehicles adopt.
In this paper, we focus on the object detection results generated by the perception system on autonomous vehicles.
The object detection function is itself one of key components for autonomous vehicles, as it allows a vehicle to account for obstacles when considering possible moving trajectories.
The object detection results of a vehicle are represented in a sensing matrix, which provides an overview of objects existing in the vehicle's surrounding environment.
Each vehicle will broadcast its sensing matrix to nearby vehicles to achieve a high-level data sharing in CAV systems.
To mitigate the potential collision of packets simultaneously transmitted by multiple vehicles, a low-latency data sharing mechanism is designed, leveraging the capture effect that is widely observed in various wireless communication techniques.
With the above design principles, we propose a \underline{L}ow-\underline{L}atency and high-\underline{L}evel (L3) data sharing protocol for connected and autonomous vehicles.

\subsection{Contributions}
Inaccurate object detection and recognition are major impediments in achieving a powerful and effective perception system on autonomous vehicles. 
To address these issues, we propose the L3 protocol in which an autonomous vehicle combines its own sensing data with that of other vehicles to help enhance its own perception. 
We believe that data redundancy, as mentioned, is the solution to this problem and we can achieve it through data sharing and combination between autonomous vehicles.
The L3 protocol is effective and efficient, which can improve the detection performance and driving experience thus providing better safety. 
Specifically, we make the following contributions.
We propose the mechanism to divide a digital map into sensing zone and vehicles will only exchange sensing data about one zone in which it current resides, leading to scalable data sharing.
The object detection results on each vehicle are represented in a sensing matrix, which facilitates a quick information sharing among vehicle, leveraging the capture effect.
The proposed L3 protocol is evaluated in simulations and it significantly outperforms existing solutions, i.e., offering a lower network latency in sharing data among vehicles.


\section{Data Sharing for Connected and Autonomous Vehicles}
To design an efficient data sharing protocol for connected autonomous vehicle, it is necessary to first understand the characteristics of data produced by various sensors on autonomous vehicles.
In this section, we investigate the mismatch between the huge amount of data generated from autonomous vehicles and the limited network bandwidth available for vehicular communications.

\subsection{Characteristics of Data on Autonomous Vehicles}
Autonomous vehicle is typically equipped with various types of sensors to obtain fine-grained, real-time and accurate information about its surrounding driving environments.
The perception system on an autonomous vehicle usually consists of several LiDAR, Radar, camera sensors, ultrasonic sensors, GPS and IMU (inertial measurement unit) sensors.
Through these sensors, sheer amount of data will be generated and these data must be processed in real time. 
Each autonomous vehicle will collect almost 4,000 Gigabyte data per day, according to~\cite{big}.
A LiDAR sensor, e.g., Velodyne LiDAR HDL-64, will generate 9.75 Mbps data when it scans at 5Hz, and up to 39 Mbps at 20Hz.

LiDAR is an essential component for autonomous vehicles as as it is used to detect dynamic and static objects including other cars or pedestrians in order to navigate around them. 
LiDAR is also applied to create high-definition maps and achieve high-accuracy localization of autonomous vehicles.
Due to the popularity of installing LiDAR sensors on autonomous vehicles, in this paper, we use the point cloud data generated by LiDAR as a case study to illustrate how the proposed L3 protocol works for connected and autonomous vehicles.

To process the data generated by LiDAR sensors, several methods are proposed, e.g., MV3D \cite{mv3d} and VoxelNet \cite{voxelnet}, to detect the objects existing in point cloud data. 
Due to the sparsity of LiDAR data, it is quite challenging to accurately detect all objects in point cloud data.
Recently, VoxelNet~\cite{voxelnet} has announced its experiments on the KITTI dataset, i.e., it offers an acceptable object detection performance on LiDAR data.
However, its detection accuracy is far from the performance of camera-based solutions.
The average car detection precision of VoxelNet is only 81.97\%.
For smaller objects, e.g., pedestrian and cyclist,
its average precision drops to 57.86\% and 67.17\%, respectively.
While in a hard condition, VoxelNet's detection accuracy of car, pedestrian and cyclist further drop to 62.85\%, 48.87\%, and 45.11\%, respectively.

\subsection{Detection Failures on Autonomous Vehicles}
Detection failures occur on autonomous vehicles for various reasons, e.g., objects are too far away, low-quality sensing data, errors in object detection algorithms. 
Therefore, it is critical to share data among autonomous vehicles to achieve cooperative perception wherein vehicles help each other to gain a better perception of their surrounding environments.
%
Leveraging the sensing data provided from other vehicles, an autonomous vehicle can essentially extent its sensing range and enhance its sensing capability, e.g, accurately detect more objects on roads.

\begin{figure}
\centering
\includegraphics[width=0.49\textwidth]{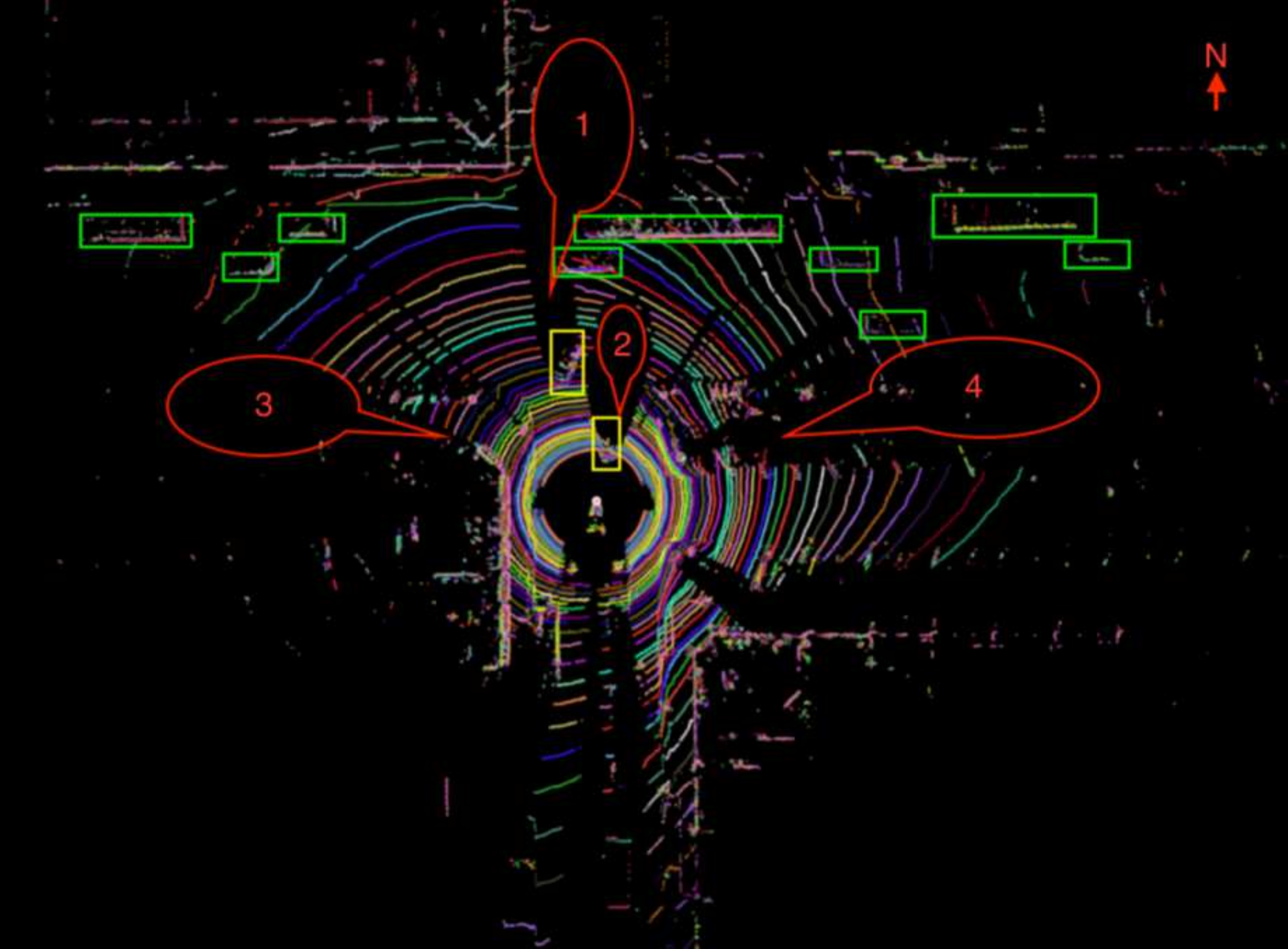}
\caption{Detection failures occur on individual vehicles because objects are blocked or exist in the blind zones of sensors.}
\label{figure:block}
\end{figure}

Using LIDAR sensor as an example, we illustrate several cases where detection failures could happen on individual vehicles that rely only on their own sensors.
As shown in Fig.~\ref{figure:block}, we collect LiDAR data on an autonomous vehicle, referred as the sensing vehicle. 
It stops in front of an intersection and move towards the north direction.
In the figure, we identify four major areas (marked with numbers) that are blocked by obstacles around the intersection.
The area 1 is totally non-observable as it is completely blocked by the vehicles moving along the west-east direction.
These vehicles are indicated by green boxes, and we can see there is a big truck blocking the sight of our autonomous vehicle.
Similar situations can be found in area 2, in which a car (marked as yellow) is located in front of the sensing vehicle.
The relatively-large objects (e.g., buildings and trees) are the root cause of the blocked areas 3 and 4.
To improve the perception capability of the sensing vehicle, it would be beneficial to let other vehicles share what they sense.
For example, the green-boxed vehicle(s) can help provide information within areas 1, 3 and 4.
The objects in area 2 could be detected by the yellow-boxed car, and the information can be shared with the sensing vehicle.

\begin{figure}
\centering

\begin{subfigure}[t]{.24\textwidth}
\centering
\includegraphics[trim=0 35 0 0, clip,width=\linewidth]{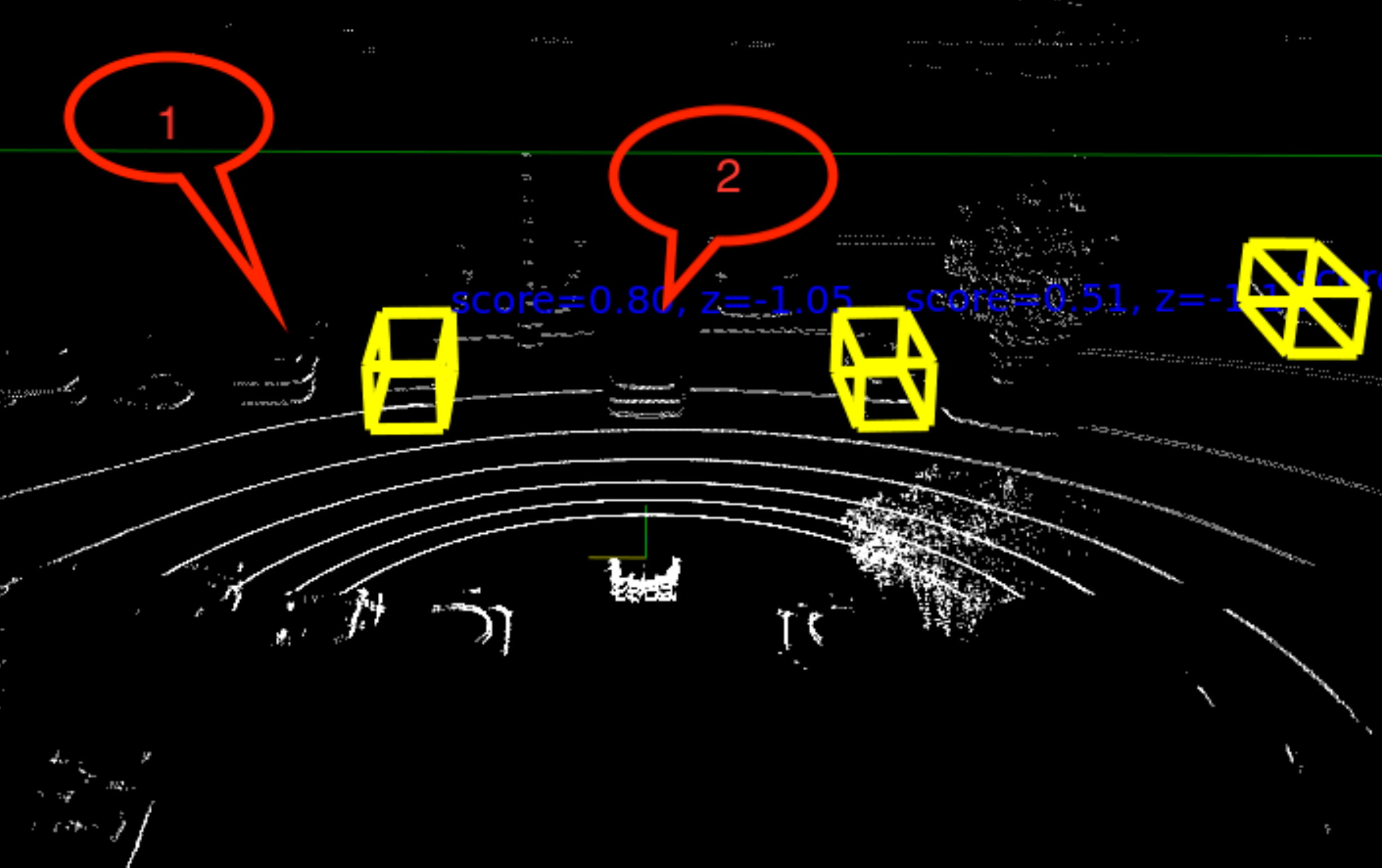}
        \caption{}
\end{subfigure}
\hspace{-5pt}
\begin{subfigure}[t]{.24\textwidth}
\centering
\includegraphics[width=\linewidth]{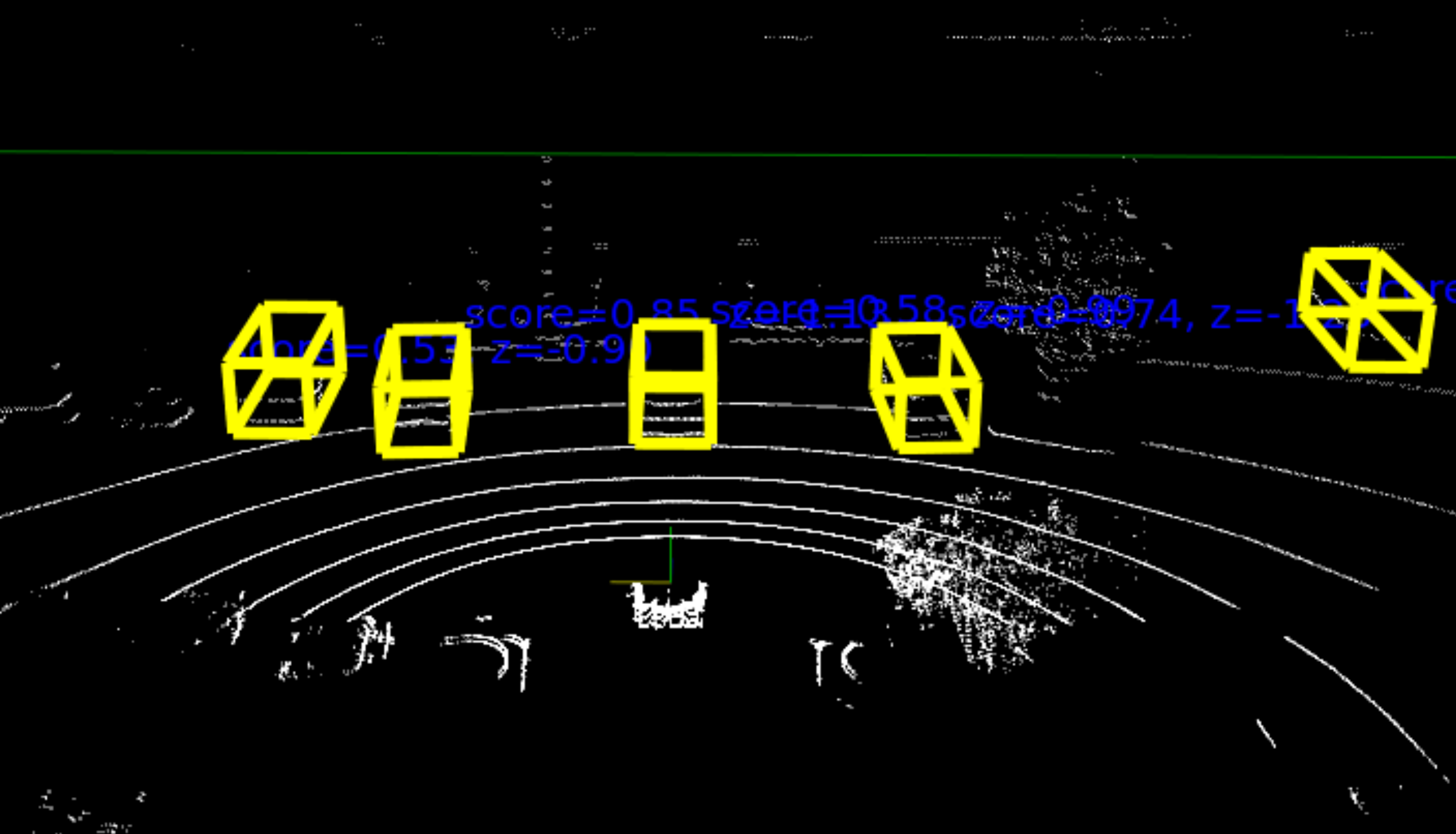}
\caption{}
\end{subfigure}

\medskip

\begin{subfigure}[t]{.24\textwidth}
\centering
\vspace{0pt}
\includegraphics[width=\linewidth]{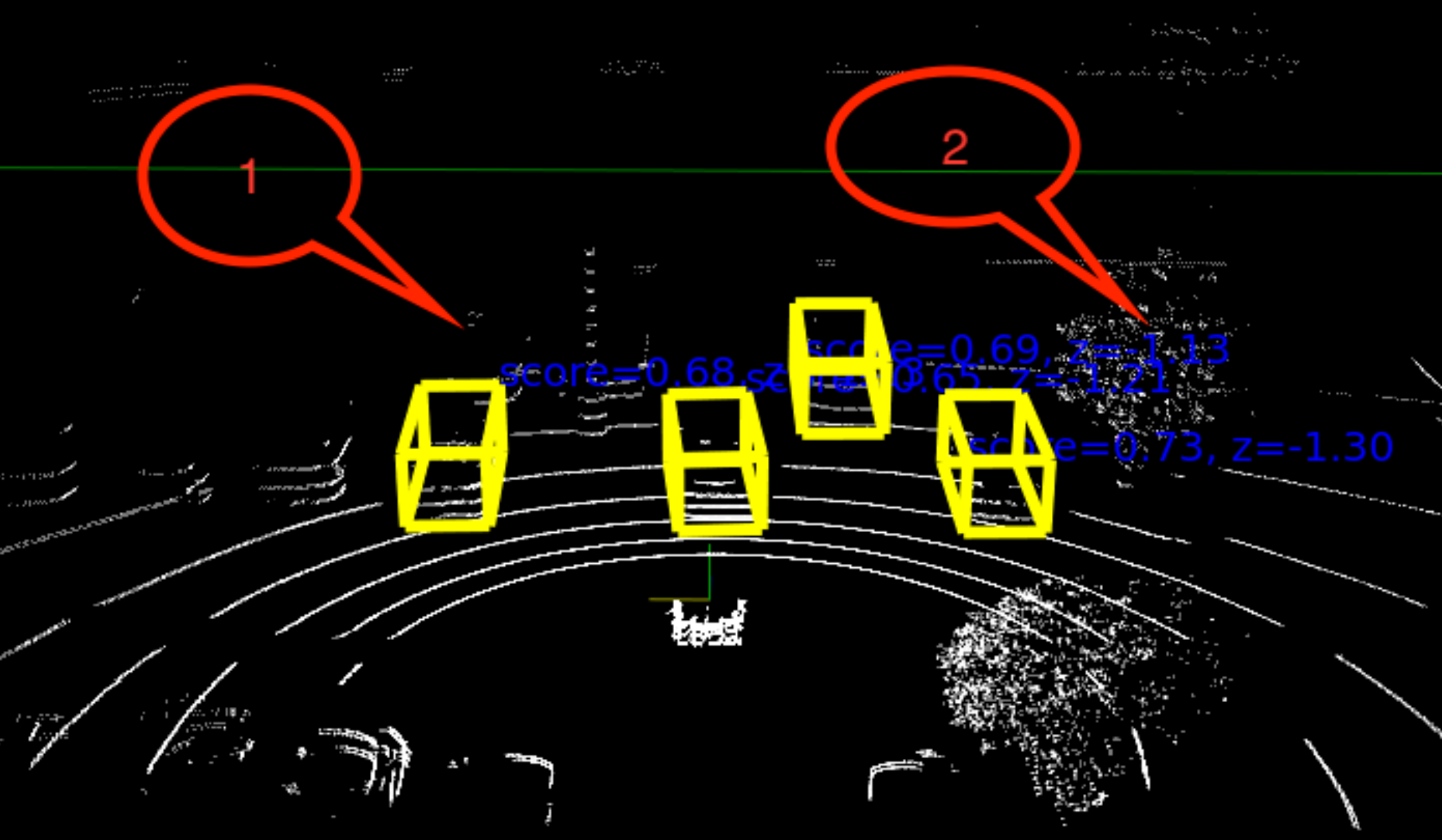}
\caption{}
\end{subfigure}
\hspace{-5pt}
\begin{subfigure}[t]{.24\textwidth}
\centering
\vspace{0pt}
\includegraphics[width=\linewidth]{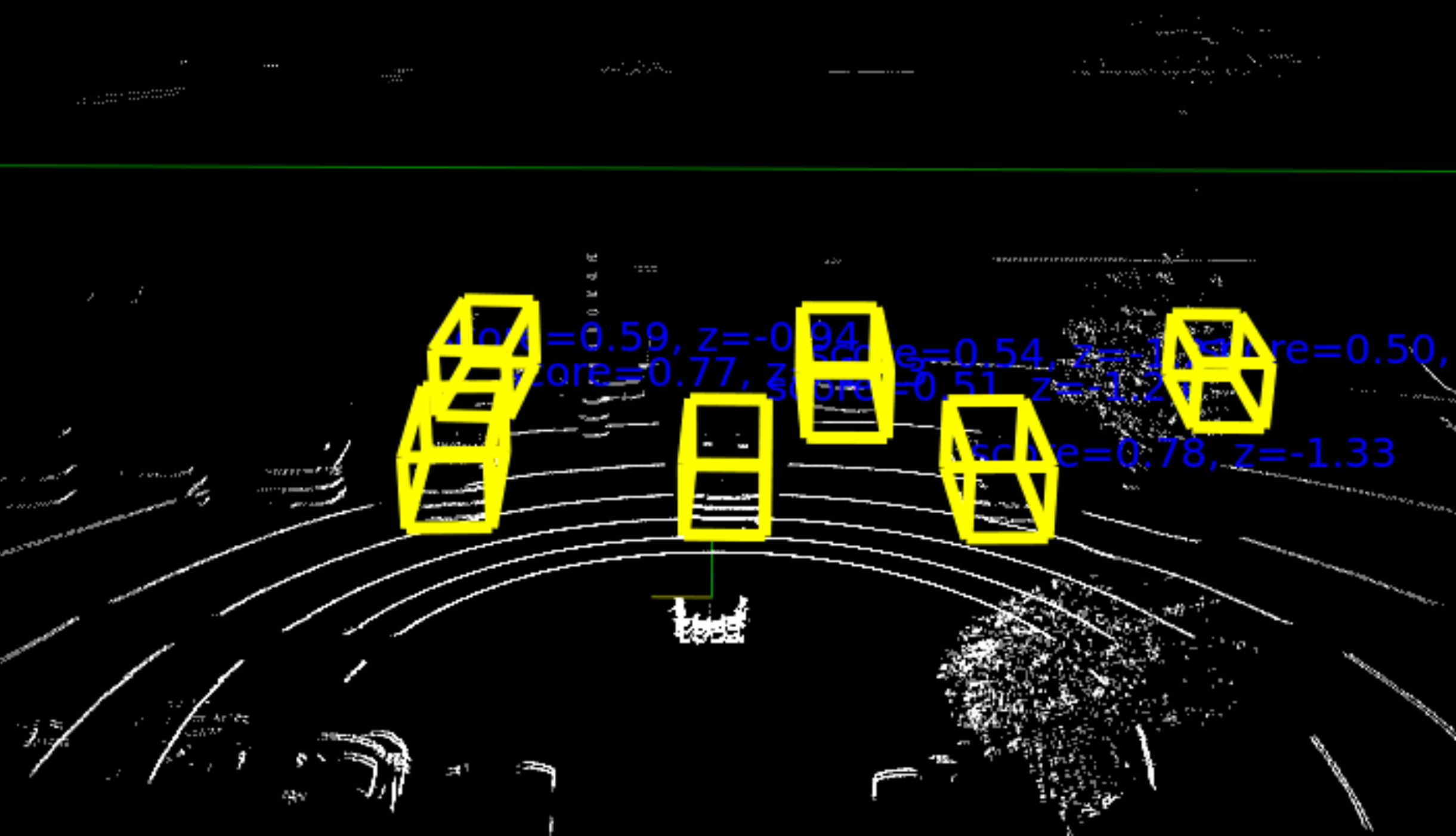}
\caption{}
\end{subfigure}

\caption{Objects detected by different vehicles at different time instances. (a) Objects (yellow boxes) detected by vehicle 1 at time $t$. (b) Objects detected by vehicle 1 at $t+5 s$. (c) Objected detected by vehicle 2 at $t$. (d) Objects detected by vehicle 2 at $t+5 s$.}
\vspace{-20pt}
\label{fig:car4}

\end{figure}

Detection failures on autonomous vehicles could also happen due to bad recognition, e.g., sensing data is too weak or is missing.
As shown in Fig.~\ref{fig:car4}, we provide several driving scenarios that are recorded in the T$\&$J dataset~\cite{cooper}.
These are the LiDAR data of an autonomous vehicle, with each key frame being a time step forward from the previous scanning position as the vehicles is moving along a straight path.
Detected objects/vehicles are marked by yellow boxes, including driving and parking vehicles. 
As we can see in Fig.~\ref{fig:car4}, in the top two frames, we have vehicle 1 detecting only 3 objects. 
When the vehicle moves forward, however, it is able to detect previous undetected objects. 
The same happens with vehicle 2, except in this case, the objects are hidden from the LiDAR sensor in the previous time step by other objects in the way.

We can imagine how dangerous this situation is should the vehicle still be blind to those regions in its path. 
Should the vehicle not detect moving objects on a collusion course with itself, then an accident is bound to happen. 
Similarly, detection failures can also contribute to the same situation should the camera sensors being obstructed in bad weather conditions.
However, this problem could be fixed if nearby vehicles exchanged sensing information.

\vspace{-10pt}

%

\subsection{Challenges in Vehicular Networks}
It is impractical to directly transmit the raw sensing data through current known wireless networks available to autonomous vehicles.
Optimally, the detected objects are labeled with detailed sensing information before transmitting out. 
The information of detected objects should include when, where and which kind of object is detected by what sensor on what vehicle, along with what size of this object and its movement conditions.
We can image it will be still challenging even if flooding the high-level object detection results in a high frequency among vehicles.

According to the research work~\cite{moveset}, tested vehicles are communicating using DSRC (Dedicated Short Range Communications) and Cellular networks along the Interstate highway I-90 in the Montana state, USA.
It is found that the DSRC throughput between two moving vehicles is less than 3 Mbps when the BPSK modulation technique is applied.
As for the cellular network performance, using Verizon and AT\&T carriers, it is shown that the LTE network can support up to 4.5 Mbps throughput, and 3G network only offer $<2$ Mbps throughput.
In summary, none of existing wireless network technology would support the high-level data sharing among autonomous vehicles.

\begin{figure}
\centering
\begin{subfigure}{0.4\textwidth}
                \includegraphics[trim=50 0 0 0, clip,width=\textwidth]{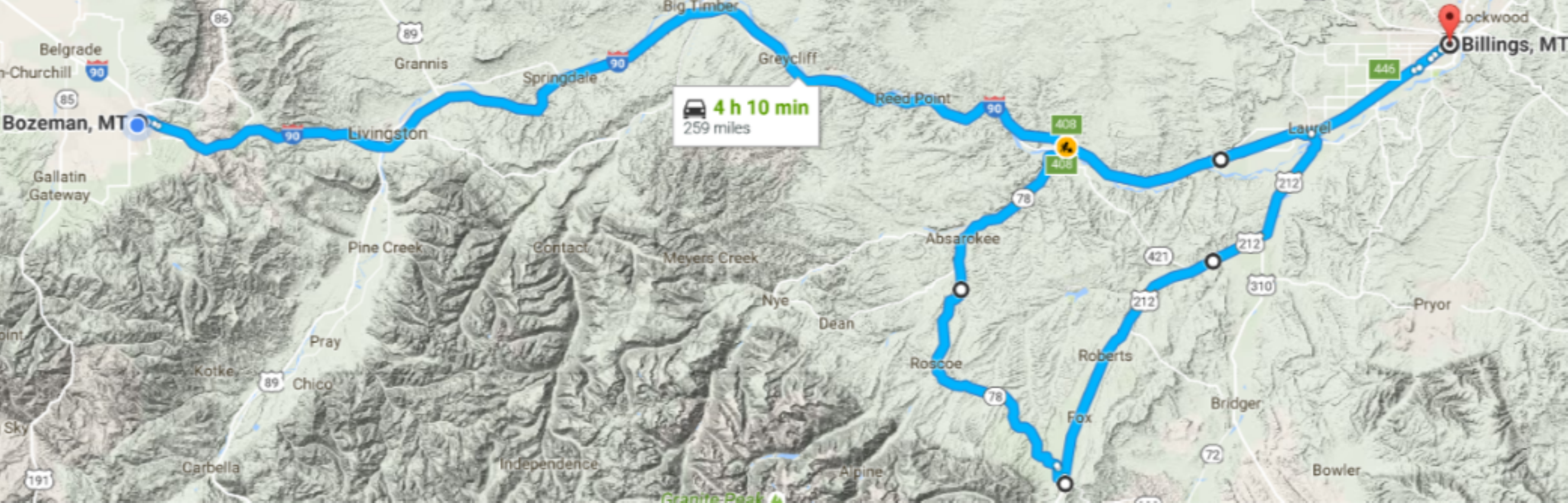}
                \caption{}
                \label{fig:1-1}
                \end{subfigure}
                
        \begin{subfigure}{0.4\textwidth}
                \includegraphics[trim=22 25 50 10, clip, width=\textwidth]{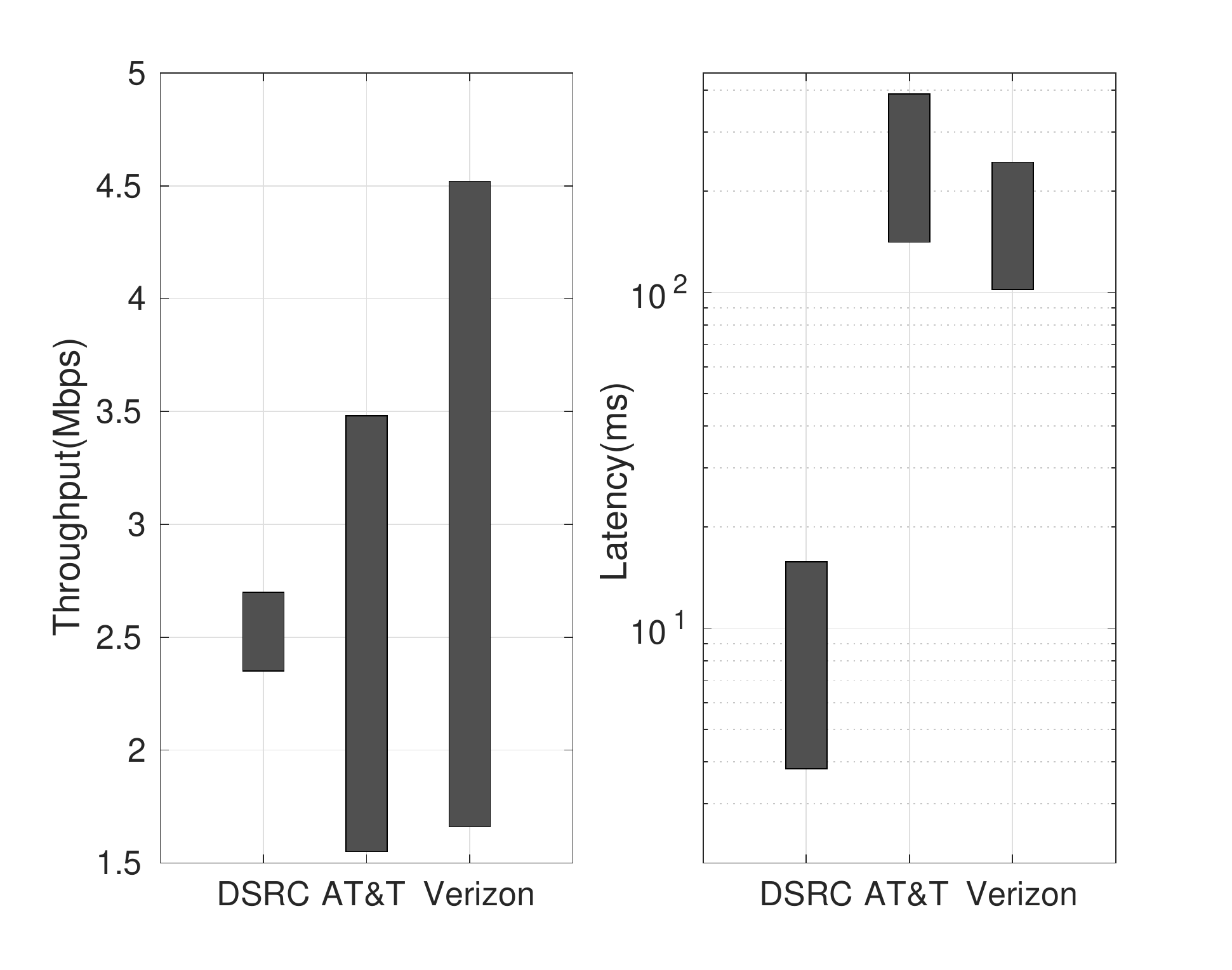}
                \caption{}
                \label{fig:thr-lat}
        \end{subfigure}
        \caption{Field tests of V2X networking technologies. (a) Measurement routs along the I-90 freeway in USA. (b) Throughput and latency of DSRC and cellular networks.}
        \vspace{-20pt}
        \label{fig:roadtest}
\end{figure}

Another technical challenge lie in vehicular networks is the large network latency introduced by transmitting the sensing data among vehicles.
For V2X communications, especially in V2V communications, low latency is required because of the high mobility of autonomous vehicles.
As shown in Fig. \ref{fig:roadtest}, hundreds of millisecond delay is observed for cellular networks, while a substantially lower delay for DSRC.

\section{\underline{L}ow-\underline{L}atency High-\underline{L}evel (L3) Data Sharing Protocol for Connected and Autonomous Vehicles}
The huge amount of data can become impractical to transmit over any existing wireless networks due to unacceptable network latency and limited, especially in a mobile environment with a large number of vehicles.
In order to develop the low-latency, high-level (L3) data sharing protocol, the high-volume of data must be reduced appropriately.
While the amount of to-be-shared data is reduced, we must guarantee the useful information obtained from the raw sensing date is still kept.
Another challenge in a CAV system is that what vehicles share may not be trustworthy~\cite{yang2015towards,cheng2019trust}, which is an important issue but out of the scope of this paper.
Therefore, we assume here that all data exchanged between vehicles are trustworthy, although detection errors may exist in the data.
%

%
\subsection{High-Level Data Sharing}
Based on the data shared from others, a vehicle must be able to extend its sensing range or increase its sensing capability; otherwise, the data transmission would be useless and should be omitted.
For example, blocked areas behind obstacles on the road could not be sensed, while this can be filled in by collecting ``unseen'' information from others.
Meanwhile, vehicles in adjacent districts or crowded areas can keep their connections for a longer duration, hence data sharing can greatly help them get more useful information. 
In summary, complementary data are always the most valuable information to share among vehicles.

\subsubsection{Sensing Zones on Digital Map}
Generally, letting every vehicle report all observations they make would provide enough information for object detection.
However, this is not the case as doing so would transmit an enormous amount of redundant data.
For example, in crowded areas, many vehicles may transmit a slight variant of the same information.
The effectiveness drops rapidly as redundancy increases.
%
To address this issue, we compress sensing results into small data packets to reduce the network traffic.

\begin{figure}
\centering
\includegraphics[trim=0 80 0 100, clip, width=0.48\textwidth]{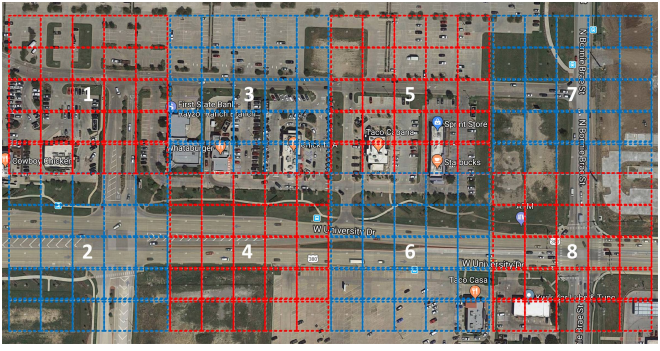}
\caption{A digital map is divided into equal-sized zones and blocks so that each vehicle is positioned into one zone and detected objects are placed into blocks.}
        \vspace{-20pt}
\label{figure:zone}
\end{figure}

We introduce an approach to position every vehicle into a zone pre-indexed on a digital map.
As shown in Fig.~\ref{figure:zone}, a digital map is divided into equal-sized zones, depicted as groups of red and blue blocks. 
Depending on the sizing choices, the sensing area of a particular sensor could be a few zones or dozens of blocks. 
For a particular vehicle, it will be located in only one zone.
Should it occupied two adjacent zones, the one it most recently touched is considered the zone where it resides in.
For this vehicle, the information of objects within its zone becomes more important than those from other zones.

\subsubsection{Sensing Blocks within A Zone}
While vehicles move on roads, each one of them will locates itself (e.g., 10Hz) into one zone based on its current location informed by its GPS sensor.
As each zone is assigned an index in the pre-installed digital map that is available to all vehicles, vehicles in the same zone would share information to each other.
When a vehicle is moving on the road, from its sensing data, it can detect various objects, including pedestrians, cars, motorcycles, and bicycles.
These objects are then labeled with their locations and size information.
The smaller the blocks, the more details about the objects,  and thus the larger communication overhead on the vehicular network.

Once a vehicle enters an indexed zone, it maps its sensing information (i.e., the object detection results) to corresponding blocks.
If a block is occupied by a detected object, the location of this blocked will be marked as object detected. 
Otherwise, there is either no object in the block or the vehicle is uncertain about whether an object exists in the block.
It worth noting that in some case a block may be out of the sensing range of a vehicle, and this case must be considered when we encode the information of each block.
To illustrate the concept of high-level sensing data sharing among autonomous vehicles, we only consider one type of objects, e.g., cars, detected by a vehicle. 
For the blocks within the zone of this particular vehicle,  four possible values will be assigned: \textit{No object}, \textit{Objected detected}, \textit{Out of sensing}, and \textit{Uncertain}.

In this way, the value of each block could be represented by two bits, denoted as $b_1b_0$.
Here, $b_1$ indicates whether the block is sensed (1) or not (0).
If $b_1  = 1$, $b_0$ indicates whether objects exist (1) or not (0).
When $b_1 = 0$ , $b_0$ presents whether this block is blocked (1) or out of sensing range (0).
In summary, we can assign 10 (No object), 11 (Objects detected), 00 (Out of sensing) and 01 (Uncertain) four possible values to each block within the sensing area of a vehicle.
\begin{figure}[!h]
        \centering
        \begin{subfigure}{0.48\textwidth}
                \includegraphics[width=\textwidth]{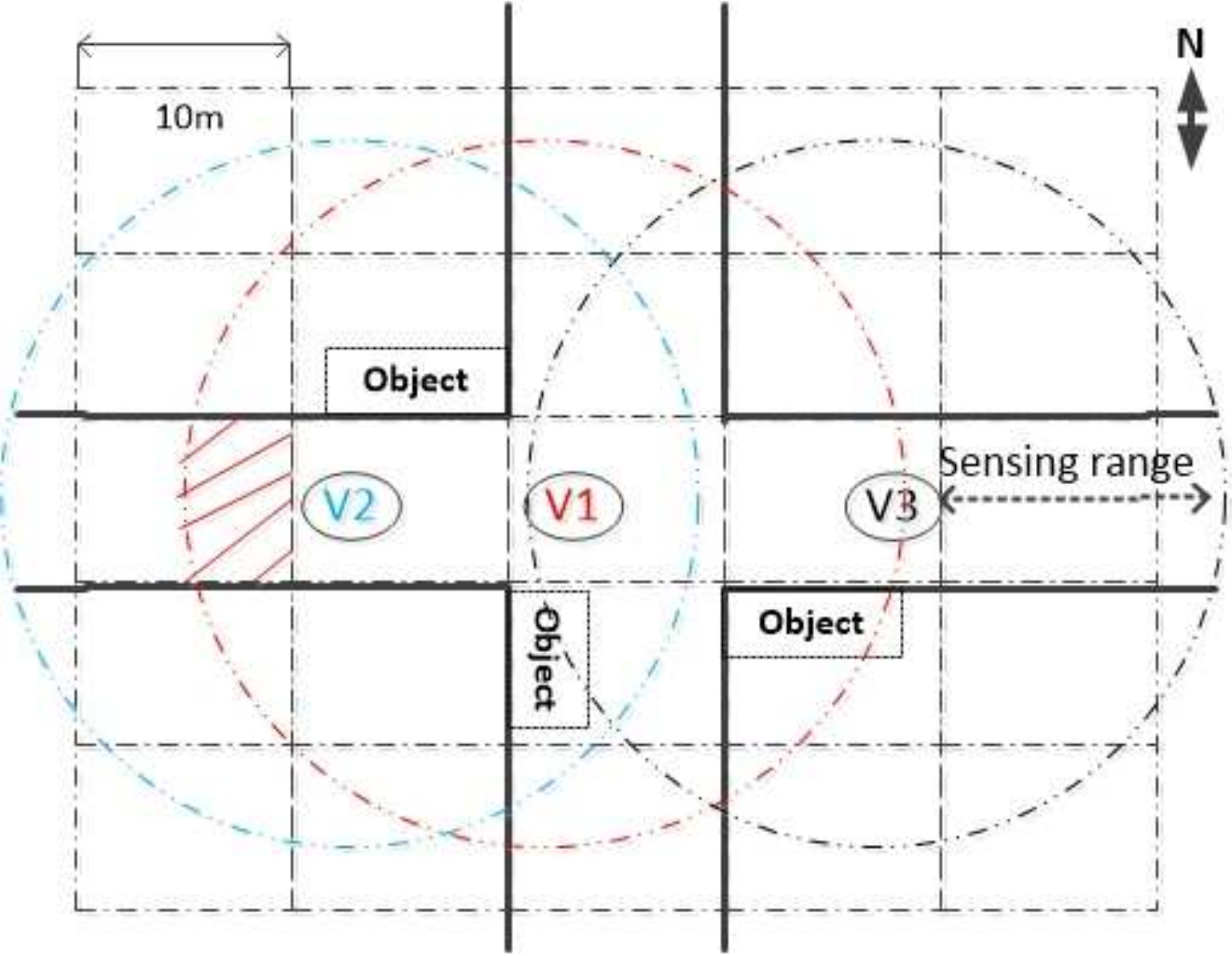}
                \caption{}
                \label{figure:road}
        \end{subfigure}
    
        \begin{subfigure}{0.5\textwidth}
                \includegraphics[width=\textwidth]{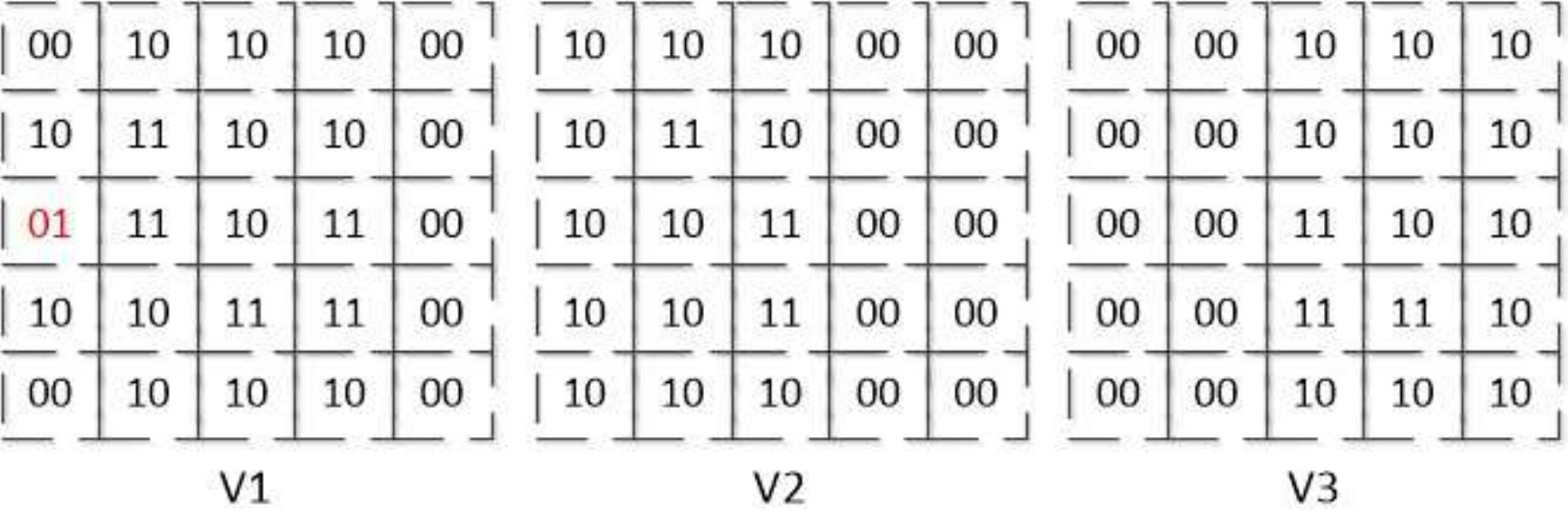}
                \caption{}
                \label{fig:3matrix}
        \end{subfigure}
        \caption{(a) Three vehicles are located within an intersection where the front view of vehicle 1 is blocked by vehicle 2. Dashed circle indicates the sensing range and the shaded area depicts the blind zone of vehicle 1. (b) Sensing matrices generated by three vehicles. The element with value of (01) in vehicle 1's sensing matrix indicates that vehicle 1 has no idea if there is any object in that block. The element with a value of (00), (11), or (10) indicates the corresponding block is out of sensing range, contains objects and no objects, respectively.}
                \vspace{-20pt}
        \label{fig:mat}
\end{figure}

As Fig.~\ref{fig:mat} shows, three vehicles are located within one zone, consisted of $5 \times 5$ blocks.
Each vehicle maps its detected objects into a block in its zone.
As such, each vehicle can prepare a $5 \times 5$ matrix, with each element representing the value of a block in vehicle's zone.

%
In this way, instead of sharing raw data, a high-level object detection results captured in a matrix could be shared among vehicles.
Different from the messages defined in the SAE J2735 standard~\cite{dsrc}, smaller packets are adopted in L3, and thus smaller network bandwidth consumption is expected. 
The SAE J2735 standard defined a DSRC message set dictionary to support interoperability among DSRC applications through the use of standardized message sets.
However, the SAE J2735 packet size is usually on the level dozens of bytes; therefore, it is not considered a light-weighted solution to data sharing among autonomous vehicles. 



\subsection{Low-Latency Data Sharing}


As many vehicles may co-locate within one zone, the information shared among nearby vehicles will become a huge load of network traffic.
In addition, the frequency of data produced by sensors is usually very high, in order to meet the real-time requirements for autonomous vehicle's applications.
Given high-frequency and huge-volume of data exchange among vehicles, network collisions are inevitable and must be carefully addressed.
In this section, we propose a low-latency data sharing protocol for V2V communications, leveraging the capture effect that widely exists in wireless communications.

After the sensing data is processed, a vehicle will create a matrix to record all objects it detects and use this matrix to determine whether it needs help from others, or it is the best vehicle to provide information for others.
For example, when a vehicle can clearly sense its surrounding environment, i.e., the value of $b_1$ of all elements in its sensing matrix is 1, it is unnecessary for this vehicle to receive or process any information shared from others.
On the other hand, if a vehicle's sensing matrix contains many elements with $b_1 = 0$, it must request helps from nearby vehicles to convert these $b_1$'s from 0 to 1.
Based on this simple principle, we design a distributed data sharing protocol for connected and autonomous vehicles.
In the L3 protocol, sensing matrices are shared among vehicles in a synchronous manner where vehicles can only send data within pre-defined slotted time intervals.
Because a vehicle's local clock is continuously synchronized with the atomic clocks on the satellites, here we assume all vehicles within a zone is well synchronized.

\subsubsection{Synchronous Data Communication based on Capture Effect}
As DSRC was standardized as the V2V communication protocol in USA~\cite{dsrc}, in this paper, we focus on designing a synchronous data transmission mechanism for DSRC.
Based on the distributed coordination function (DCF) defined in the IEEE 802.11p protocol, multiple access control is implemented based on the well-known carrier sense multiple access with collision avoidance (CSMA/CA) mechanism~\cite{csmaca}. 
The DCF approach is proved to be efficient for relatively-low network traffics, however, its performance degrades significantly in the cases where large amount of devices transmitting data simultaneously. 
In these cases, as packet collisions occurs frequently, an larger contention window on each vehicle is expected, which will not only increase the network delay but also reduce the overall network throughput.

To address the above-mentioned issues, we propose to leverage the capture effect that was widely studied in IEEE 802.11 protocols~\cite{chaos}.
Capture effect enables a receiver to correctly decode a packet when the received signal is about 3 dB stronger than the sum of the received signals from all others~\cite{capacity, design}.
As such, given multiple simultaneous wireless transmissions, only the one with the strongest received signal can be received and decoded.
To ensure capture effect, the strongest signal must arrive no later than the air time of synchronization header, after the first weaker signal~\cite{chaos}.
If these conditions are satisfied, collided packets (from the strongest signal) can be successfully decoded on the receiver.
Due to the capture effect, vehicles can receive packets despite interference from other vehicles that are transmitting packets at the same time.
As such, the network throughput is improved and the network latency is reduced.

The synchronous data communication protocol works as follows.
Vehicles owning uncertain blocks initiate the data communication process by sending their sensing matrices to nearby vehicles.
The nearby vehicles overhearing the data will receive these packets with a high probability, due to the capture effect.
On reception of these packets, the vehicles combine their own sensing data with the received ones and update their sensing matrices accordingly.
The updated sensing matrices will be again shared with other vehicles.
This data aggregation process continues in a fully distributed manner until all vehicles in the same zone have the same sensing matrix.
When the protocol is executed multiple times, several vehicles may have the same sensing matrix.
In this case, when these vehicles simultaneously send the same sensing matrix to others, a constructive inference could be observed.
Constructive inference occurs only when packets are identical and overlap with each other within 0.5 $\mu$s.
Apparently, constructive inference would speed up the data sharing process among vehicles.

\begin{figure}[h]
        \centering
        \begin{subfigure}{0.35\textwidth}
                \includegraphics[width=\textwidth]{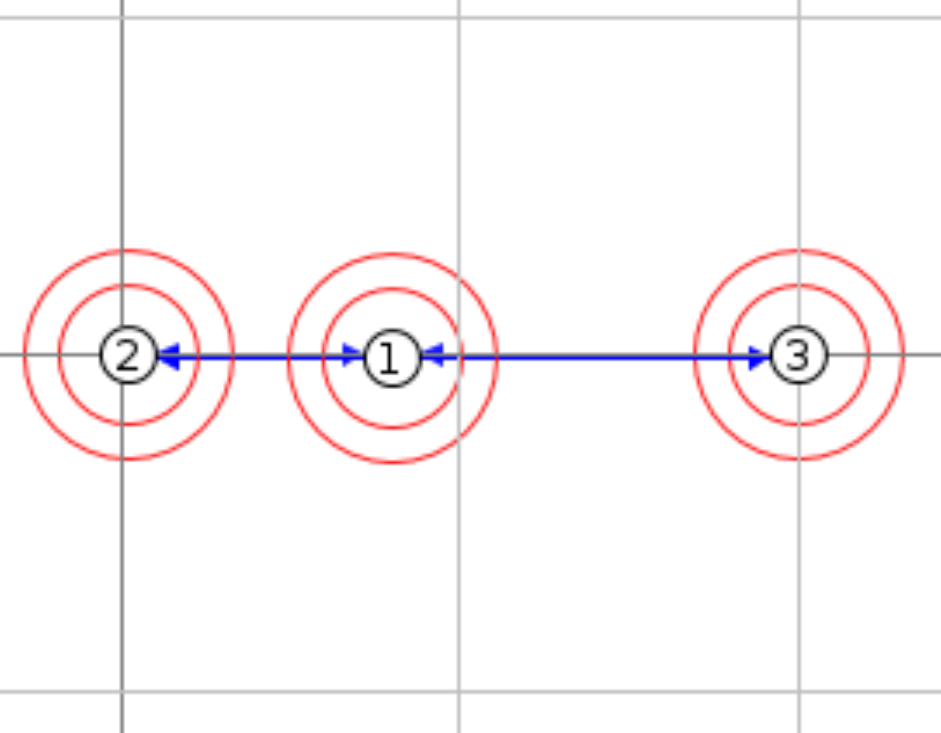}
                \caption{}
                \label{figure:3nodes}
                \end{subfigure}
        \begin{subfigure}{0.48\textwidth}
                \includegraphics[width=\textwidth]{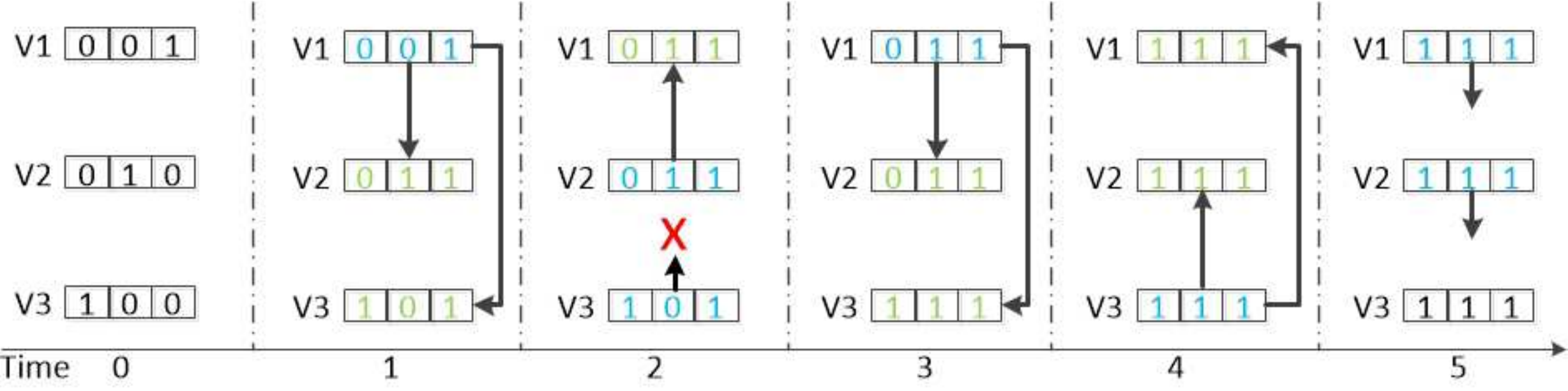}
                \caption{}
                \label{fig:3vehicles}
                \end{subfigure}
        \begin{subfigure}{0.4\textwidth}
                \includegraphics[width=\textwidth]{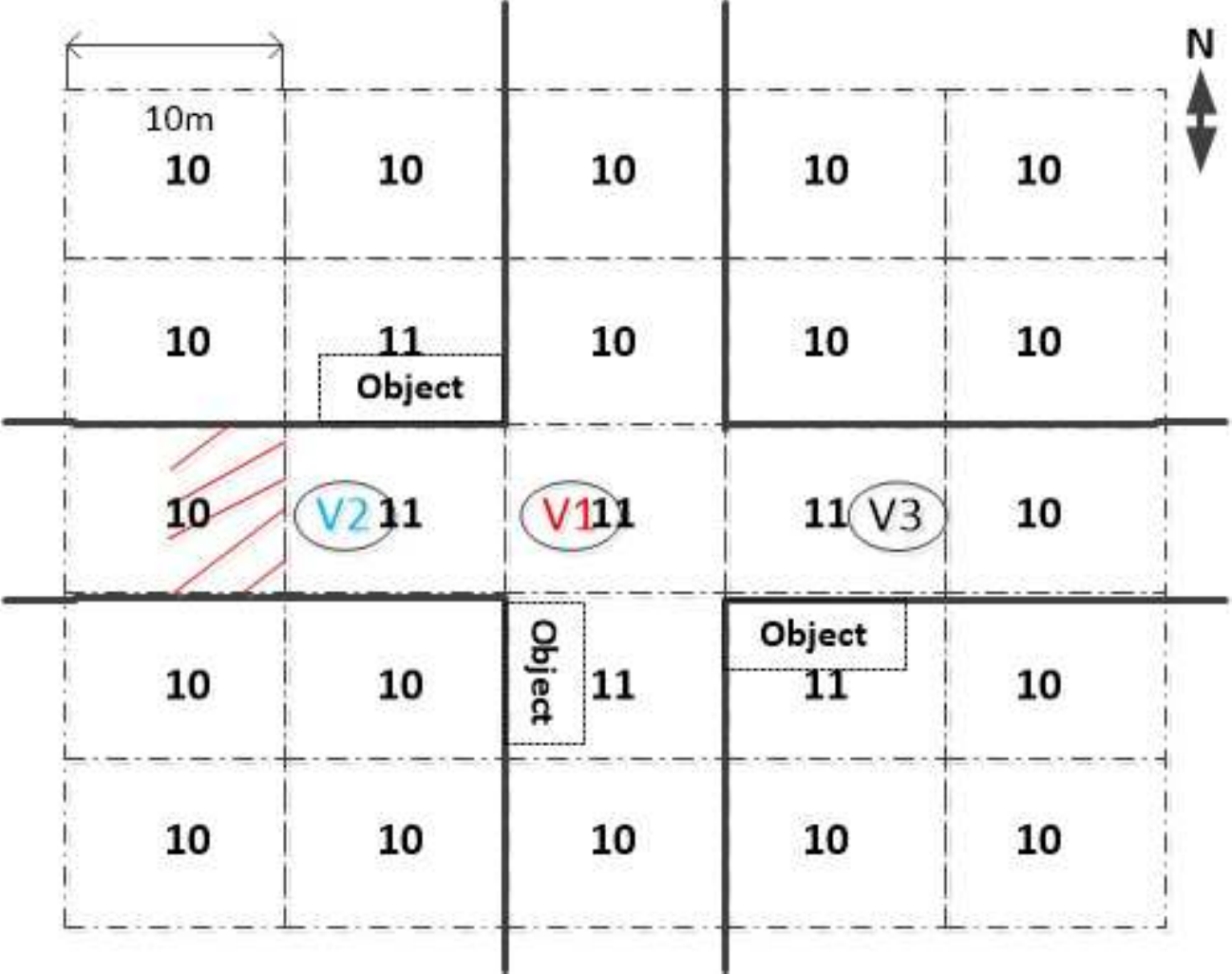}
                \caption{}
                \label{fig:col}
                \end{subfigure}
        \caption{Illustration of how sensing matrices are exchanged among three vehicles. (a) Three vehicles are located along a line with vehicle 2 being closer to vehicle 1 than vehicle 3. (b) The data communication process among three vehicles. Arrowed line indicates a packet is successfully delivered, due to capture effect. Red cross means a packet is dropped. (c) The final sensing matrix on vehicle 1. It starts the data sharing process as it has a block with a value of (01). After 5 rounds of data exchange with others, the value of this block is updated to (10). }
                \vspace{-20pt}
        \label{fig:comm}
\end{figure}

We use an example shown in Fig.~\ref{figure:3nodes} to illustrate how capture effect would facilitate faster data sharing among three vehicles.
Here, we assume vehicles 1, 2, and 3 are within the  communication range of each other.
The three vehicles are assumed to reside in a zone containing 25 $10m \times 10m$ blocks.
Based on its sensors, each vehicle could prepare a sensing matrix. 
In the example, as vehicle 2 blocks the front view of vehicle 1, there is an uncertain block in vehicle 1's sensing matrix.
According to the L3 protocol, vehicle 1 will initiate the data sharing process via sending its sensing matrix in time slot 1. 
When vehicle 2 and 3 receive the message from vehicle 1, they will aggregate the received data with their own data and update their sensing matrices.
The updated sensing matrices are then transmitted from vehicles 2 and 3 simultaneously in time slot 2.
Vehicle 1 will receive the update sensing matrix from vehicle 2, due to capture effect.
With the new information contained in the receive message, vehicle 1 will update its sensing matrix and send it in time slot 3. 
As now new information is received, vehicle 2 does not send anything in time slot 4.
At time slot 5, because all vehicles have the same sensing matrix, the data sharing process stops.

\subsection{Data Aggregation Process}
When sensing matrices are received from other vehicles, it is necessary to design a data aggregation process to combine the received data with the current one.
As we are focusing on enhancing the object detection capability of autonomous vehicles, the data aggregation must produce a sensing matrix that contains all the objects detected by the sharing vehicles.

The data aggregation process on a vehicle starts from identifying whether the received data is generated from another vehicle within the same zone.
This can be done by comparing the index of the zones where the vehicles reside.
If the received sensing data, denoted as $R_{m \times n}$, are for the same sensing zone, the aggregation will be carried out as follows.
Here, we assume there are $m \times n$ blocks within the current zone.
Similarly, we use $C_{m \times n}$ to denote the sensing matrix on the current vehicle which takes $R_{m \times n}$ to update its own sensing matrix.

To aggregate two matrices $R_{m \times n}$ and $C_{m \times n}$, we will compare all elements from these two matrices one by one.
For a particular pair of elements, we use $b^r_1 b^r_0$ and $b^c_1 b^c_0$ to represent the sensing data in the received and current sensing matrices, respectively.
If $b^r_1 = 0$, it implies the received data do not contain any useful information for the corresponding block.
Therefore, the value of $b^c_1 b^c_0$ will be kept unchanged.
On the other hand, if $b^r_1 = 1$ and $b^c_1 = 0$, the the value of $b^c_1 b^c_0$ will be replaced by $b^r_1 b^r_0$.
If $b^r_1 = b^c_1 = 1$ but $b^c_0 \neq b^r_0$, it means there is inconsistency on the object detection from the two vehicles.
In this case, as it is difficult to determine which one offers the best detection result, we consider uncertain observations were made.
As a result, the value of $b^c_1 b^c_0$ becomes $01$, which will again trigger the data sharing process.
We believe this case is very rare and only occurs occasionally. %
The aggregation process will be applied to all pairs of elements from two sensing matrices.
The entire data aggregation process is summarized in Algorithm 1.

\begin{algorithm}[h]
\SetAlgoLined
\KwResult{Sensing matrix $C_{m \times n}$ is updated.}
  \For{$\forall b^r_1 b^r_0 \in R_{m \times n}$}{
   $b^c_1 b^c_0 \leftarrow$ Corresponding element in $C_{m \times n}$\;
    \If{$b^r_1 = 1 \wedge  b^c_1 = 0$}{
     {$b^c_1 b^c_0 \leftarrow b^r_1 b^r_0$\;}
        \ElseIf{$b^r_1 = b^c_1 = 1 \wedge  b^c_0 \neq b^r_0$}
         {$b^c_1 b^c_0 \leftarrow 01$\;}
     }
    }
\label{alg:alg1}
\caption{Data Aggregation}
\end{algorithm}


Fig~\ref{fig:3vehicles} illustrates how messages are exchanged among the three vehicles.
We can see that the blocked area from vehicle 1's perspective is updated based on the sensing data provided by vehicle 2.
Such information is then transmitted to vehicle 3.
After five rounds of communications, the sensing matrix converges to the one shown in Fig.~\ref{fig:col}.
As such, vehicle 1 is able to extend its sensing capability by detecting there is no object existing in the area blocked by vehicle 2.
With the proposed L3 protocol, consistent sensing results could be derived on individual vehicles which shared their own sensing data to others.


\section{Evaluation and Result Analysis}
In this section, we evaluate the performance of L3 protocol in simulations.
Although the capture effect is widely observed on IEEE 802.11 devices, it is not yet implemented in IEEE 802.11p.
As most DSRC devices are not open-source platforms, it is prohibitively expensive to conduct reverse engineering on these devices to implement capture effect.
As such, we adopt the COOJA~\cite{cooja} simulator to evaluate L3 protocol.
Although COOJA is a contiki IEEE 802.15.4 network simulator, it can adequately approximate the data communication process among vehicles using DSRC.
With the COOJA simulator, we simulate scenarios where several vehicles communicate with each other to share the object detection results via the L3 protocol.
Particularly, we are interested in how many rounds of data communications are needed to realize a consistent sensing matrix on all participating vehicles.
Next, we use the NS-3 simulator~\cite{ns3} to simulate and measure the network delay and scalability of the L3 protocol.

\subsection{Simulation Setup}
According to the DSRC protocol, we note that the reliable communication range of DSRC is about hundreds of meters.
On the other hand, the effective sensing ranges of regular sensors, e.g., LiDAR, Radar and camera, are far less than the communication range.
In the simulations, we set a zone as a 100*100$m^2$ square, and we assume all vehicles within a zone can communicate to each other.
Meanwhile, we set each vehicle's sensing range as 25 meters.
We also set the block size as 5*5$m^2$.
As such, there will be 400 blocks in one sensing zone, i.e., 800 bits are needed to record the sensing results of each block in a zone.
As there are only 800 bits in each packet, the payload of the L3 protocol is 100 bytes. 
Overall, the simulation setup parameters are shown in Table~\ref{setting_table}.

\begin{table}
\centering
\caption {Simulation setup parameters}

\begin{tabular} {|l | c|}
\hline

\hline
Communication Range & $100m$ \\ \hline
Sensing Range & $25m$ \\ \hline
Payload Size & $100$ bytes \\ \hline
Zone Size& $100m \times 100m$ \\ \hline
Block Size & $5m \times 5m$ \\ \hline
\end{tabular}

\label{setting_table}
\end{table}

\subsection{Convergence Time}
L3 is designed to realize low-latency data sharing among autonomous vehicles, therefore, it is important to evaluate how long it takes to ensure all participating vehicles have the consistent sensing matrix.
The latency can be measured in two dimensions: (1) number of time slots taken and (2) the actual time taken to reach a consistent sensing matrix on vehicles. 
In this section, we evaluate the how many time slots are needed to complete the data sharing process among vehicles.

\begin{figure}
\centering

\begin{subfigure}{.22\textwidth}
\centering
\includegraphics[width=\linewidth]{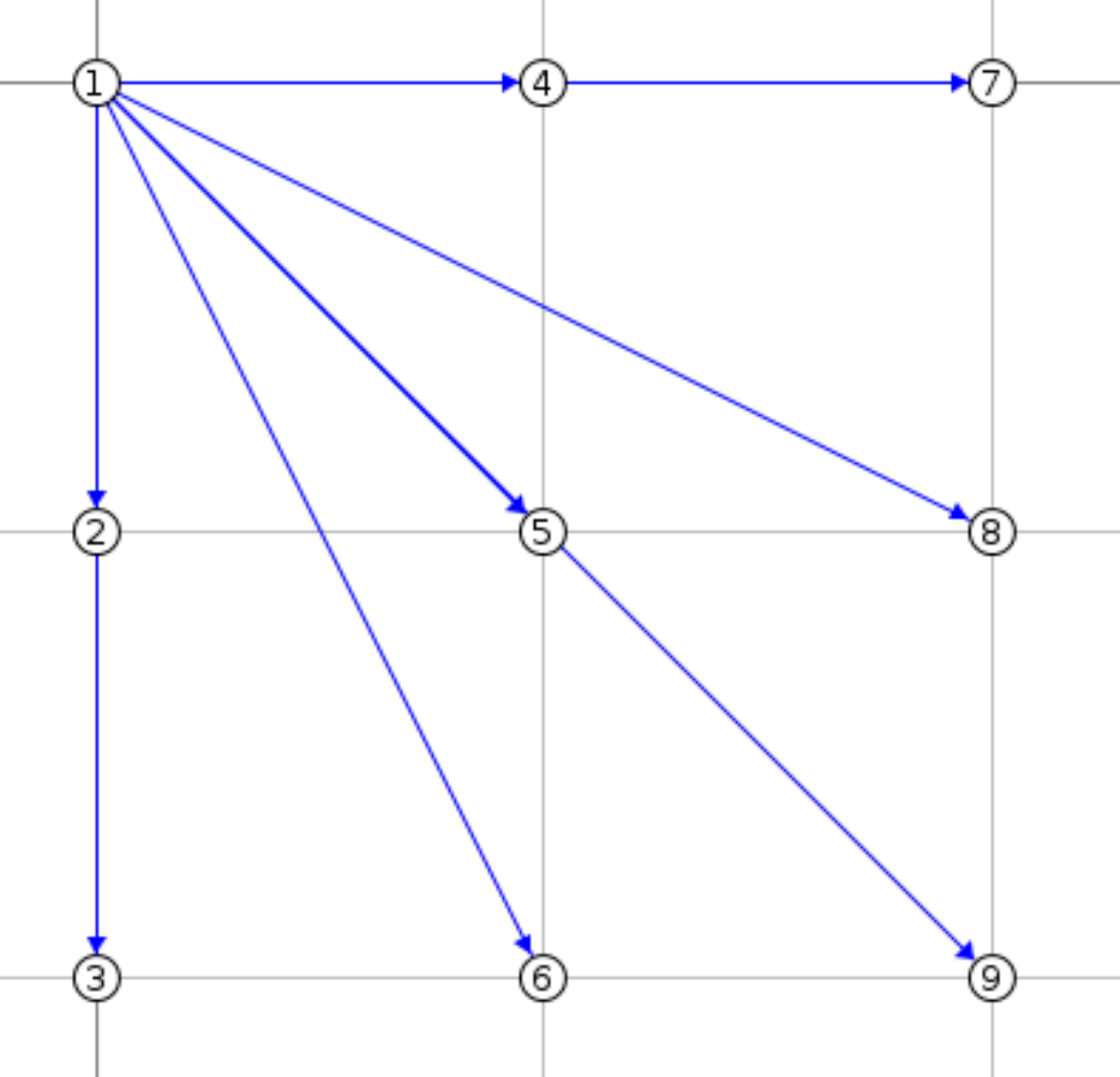}
        \caption{}\label{fig:fig_a}
\end{subfigure}
\hspace{-1pt}
\begin{subfigure}{.22\textwidth}
\centering
\vspace{0pt}
\includegraphics[width=\linewidth]{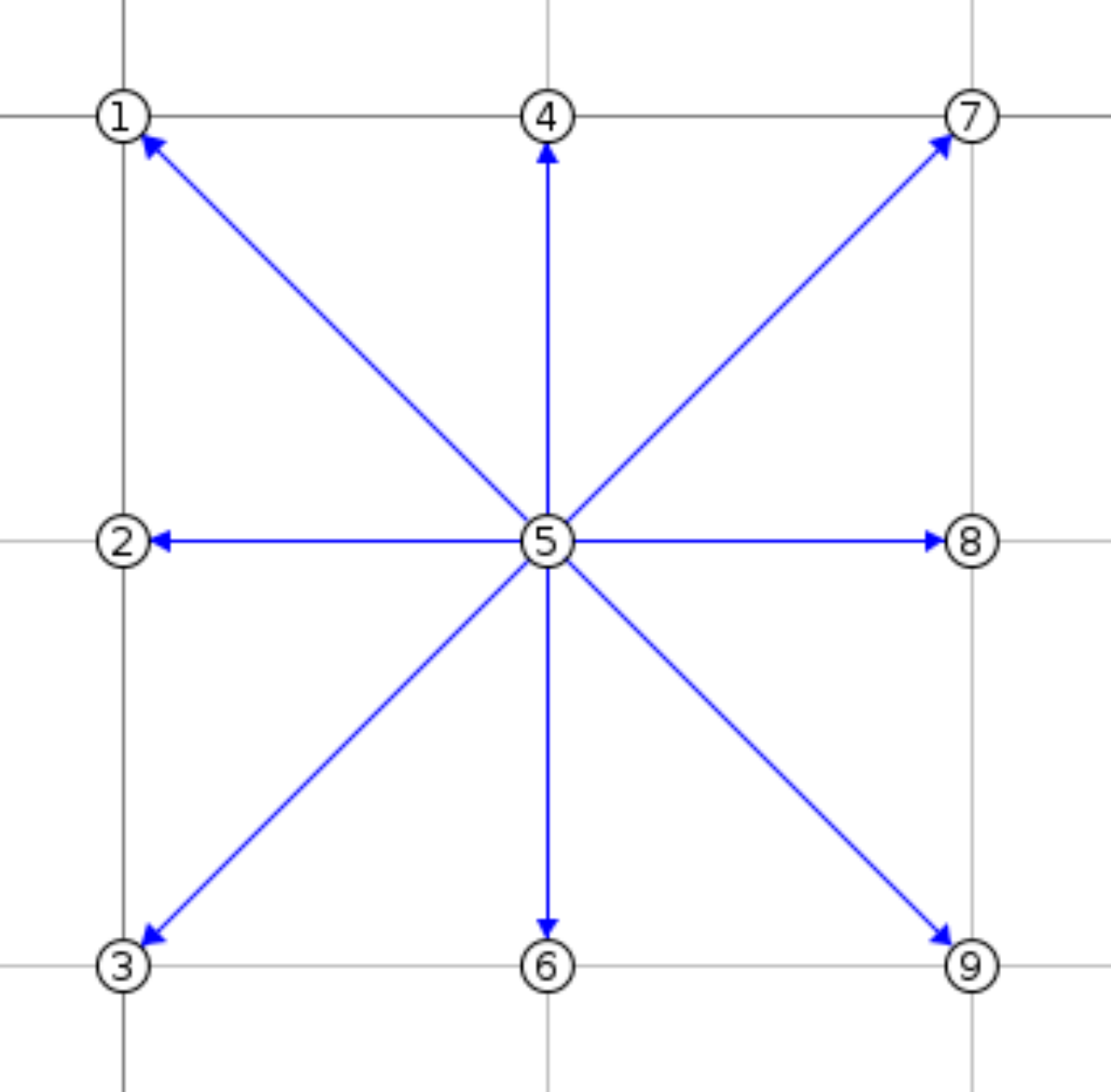}
\caption{}\label{fig:fig_c}
\end{subfigure}

\begin{subfigure}{.22\textwidth}
\centering
\includegraphics[width=\linewidth]{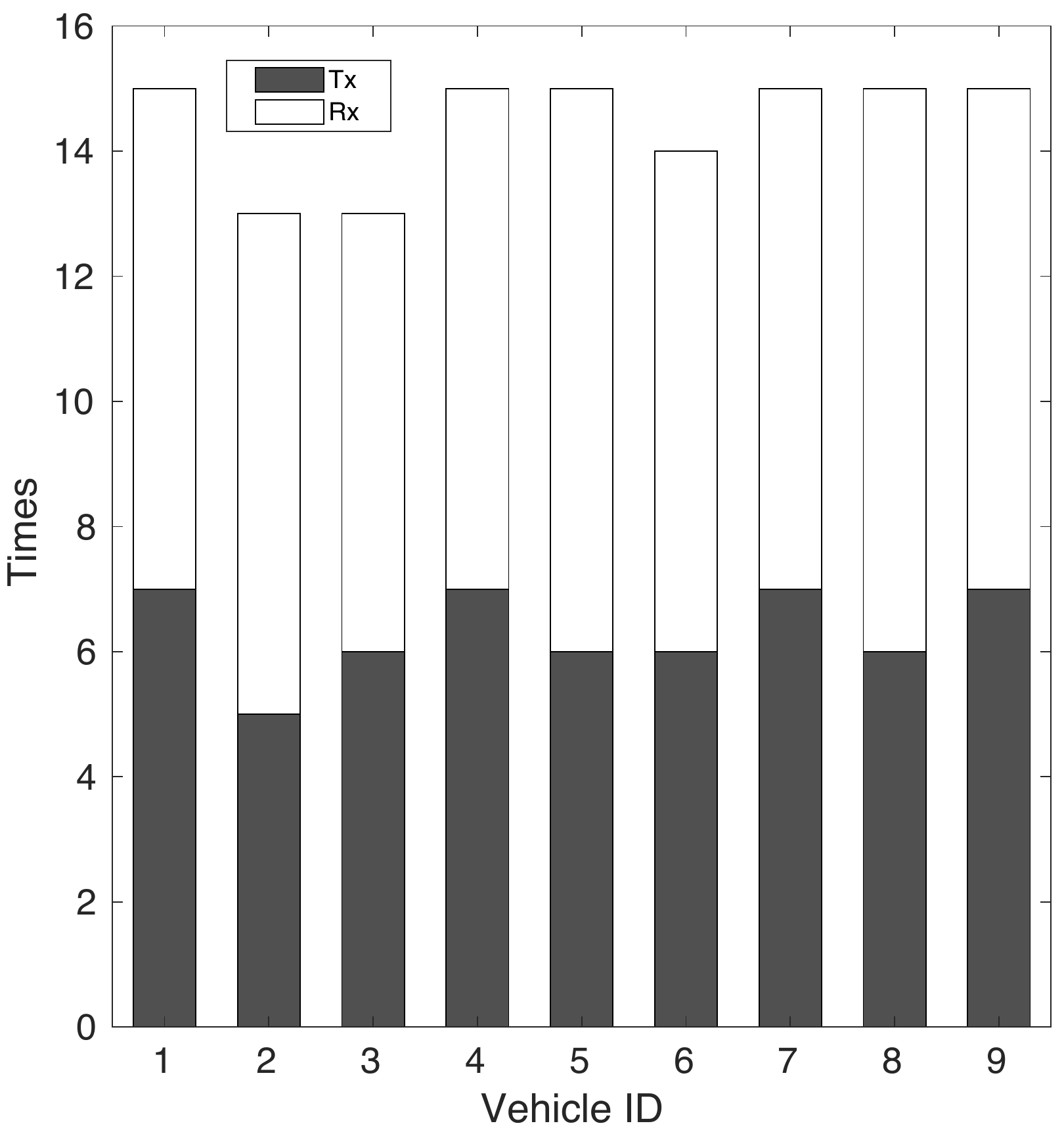}
\caption{}\label{fig:fig_b}
\end{subfigure}
\hspace{-5pt}
\begin{subfigure}{.23\textwidth}
\centering
\vspace{0pt}
\includegraphics[width=\linewidth]{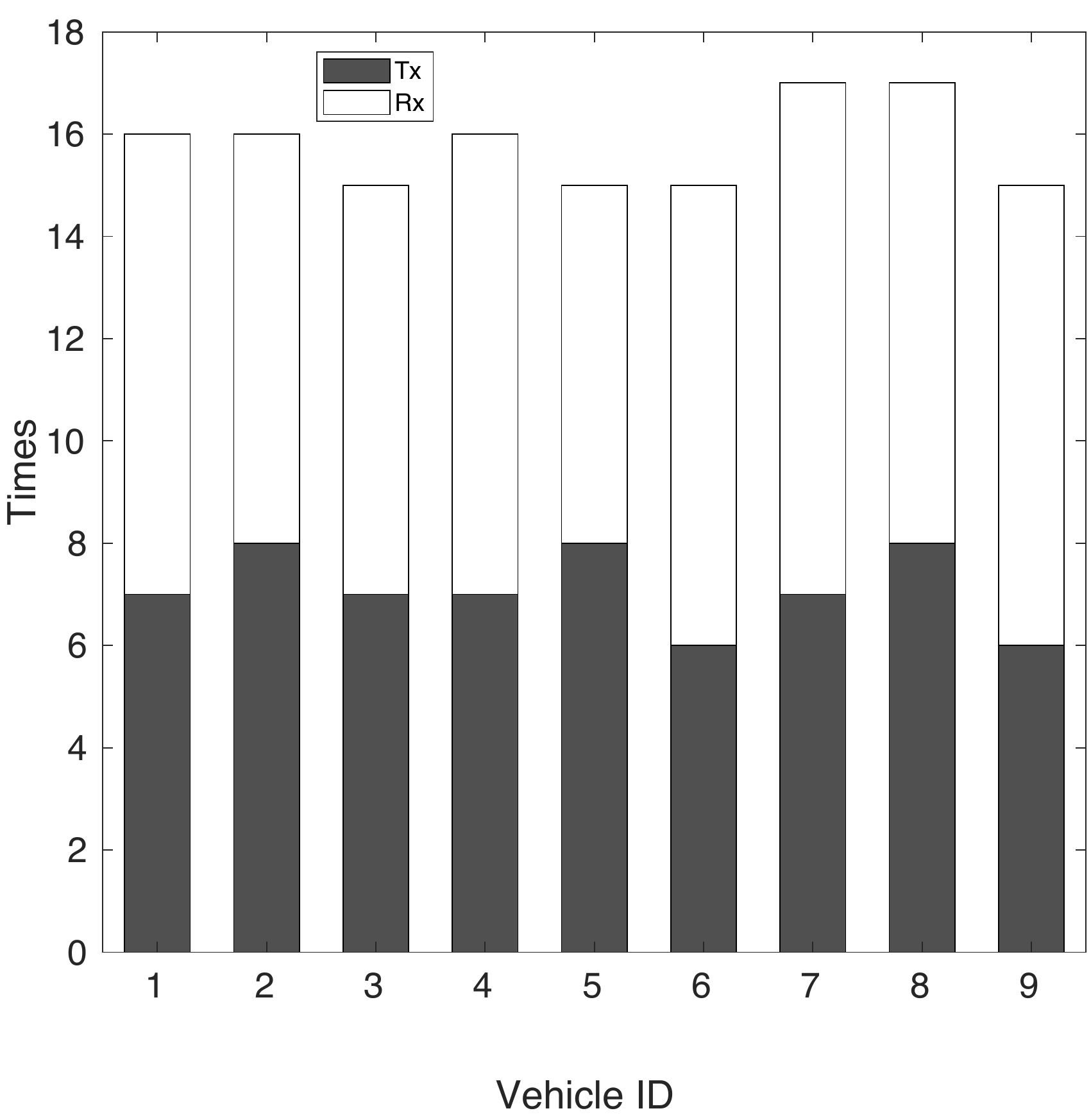}
\caption{}\label{fig:fig_d}
\label{fig:9veh}
\end{subfigure}

\medskip

\begin{minipage}[t]{.5\textwidth}
\caption{Convergence time of 9 vehicles exchanging sensing matrices between each other. (a) Vehicle 1 located at top-left corner initiates the data sharing process. (b) Vehicle 5 located in the middle initiates the data sharing process. (c) The distribution of time slots in which vehicles are in transmission and reception modes, respectively. (d) The distribution of time slots in which vehicles are in transmission and reception modes when vehicle 5 is the initiator.}
\vspace{-20pt}
\end{minipage}

\end{figure}

In the simulation, we place 9 vehicles in a grid using the COOJA simulator, as shown in Fig.~\ref{fig:fig_a}.
The horizontal/vertical distance between two adjacent vehicles is set to be 10m.
We first let vehicle 1 to initiate the data sharing process, which represents the cases where vehicles located around the corners of the grid to start communications.
We then record how many time slots a vehicle is in its transmission or reception modes, until all 9 vehicles have the same sensing matrix.
As shown in Fig.~\ref{fig:fig_b}, after a total of 15 rounds of data exchange, all vehicles reach the same sensing matrix, i.e., the sensing results converge.
For vehicle 1, it transmits its original (or updated) sensing matrix for 7 time slots and receive data from others in 8 time slots.
For vehicles 2 and 3, as they are in the perfect location of receiving sensing data, their sensing matrix converges after the 12$th$ round of data exchange.
After that, they simply broadcast the converged sensing matrix one more time.
Due to constructive interference, their transmissions will not collide even though two packets are transmitted during the same time slot.
After the 13$th$ time slot, vehicles 2 and 3 finish the data sharing process and they do not send or receive any new data.
At the 14$th$ time slot, vehicle 6 receives the message from vehicle 3, due to capture effect as vehicle 6 is closer to vehicle 3.
As such, vehicle 3 received the converged sensing matrix and concludes its data sharing process as well.
The remaining vehicles will finish the sensing matrix updating process in a similar manner.
After 15 rounds of data exchange, all vehicles obtain the same sensing matrix for the targeted sensing zone.

Next, we make vehicle 5 serve as the initiator, which represents the cases where vehicles in the middle of a sensing zone starts the data sharing process.  
As shown in Fig.~\ref{fig:fig_c}, vehicle 5 starts sending and collecting sensing data from others.
Different from our expectation, in this case, it takes a total of 17 time slots to finish the data sharing process.
This is mainly because it takes a longer time for data from vehicles at one side to propagate to those at the other side.
The distribution of transmission and reception activities of each vehicle is plotted in Fig.~\ref{fig:fig_d}.
As we can see, vehicle 3, 5, 6 and 9 receive the converged sensing matrix after the 14$th$ time slot.
They broadcast the sensing result one more time and then keep silent.
Vehicles 1, 2, and 4, on the other hand, receive vehicle 5's message at the 15$th$ time slot, due to capture effect.
As the message contains the final sensing matrix, they all halt the sharing process after one more round of broadcasting. 
The last two vehicles (7 and 8) complete their updating process at the 17$th$ time slot, and the entire sharing process is finished.

\begin{figure}
\centering
\includegraphics[width=0.45\textwidth]{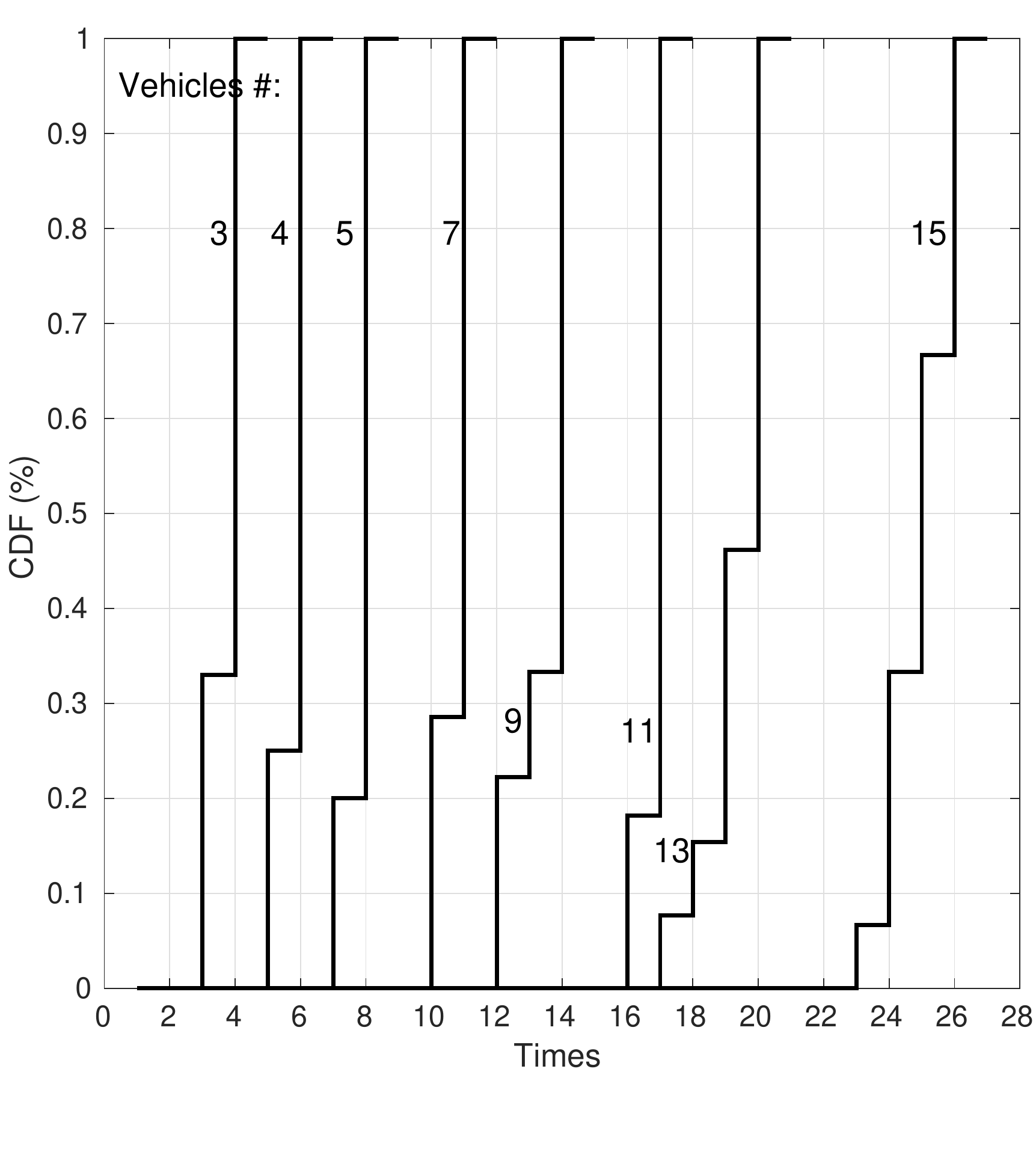}
\vspace{-10pt}
\caption{Completion times of data sharing with different number of vehicles in the network.}
\label{figure:cdf}
\vspace{-20pt}
\end{figure}

Besides the above simulations, we also conduct experiments with different number of vehicles that are randomly deployed in a certain area.
The convergence times of different scenarios, i.e., deploying 3 to 15 vehicles randomly in a zone, are summarized and plotted in Fig.~\ref{figure:cdf}.
We first deploy 3 vehicles in a zone and it takes 4 time slots to finish the data sharing among the three vehicles.
The convergence time grows as more and more vehicles join in the data sharing process.
The total number of rounds increases up to 26 time slots when there are 15 vehicles in the network.

\subsection{Network Latency}
\begin{figure}
\centering
\includegraphics[width=0.45\textwidth]{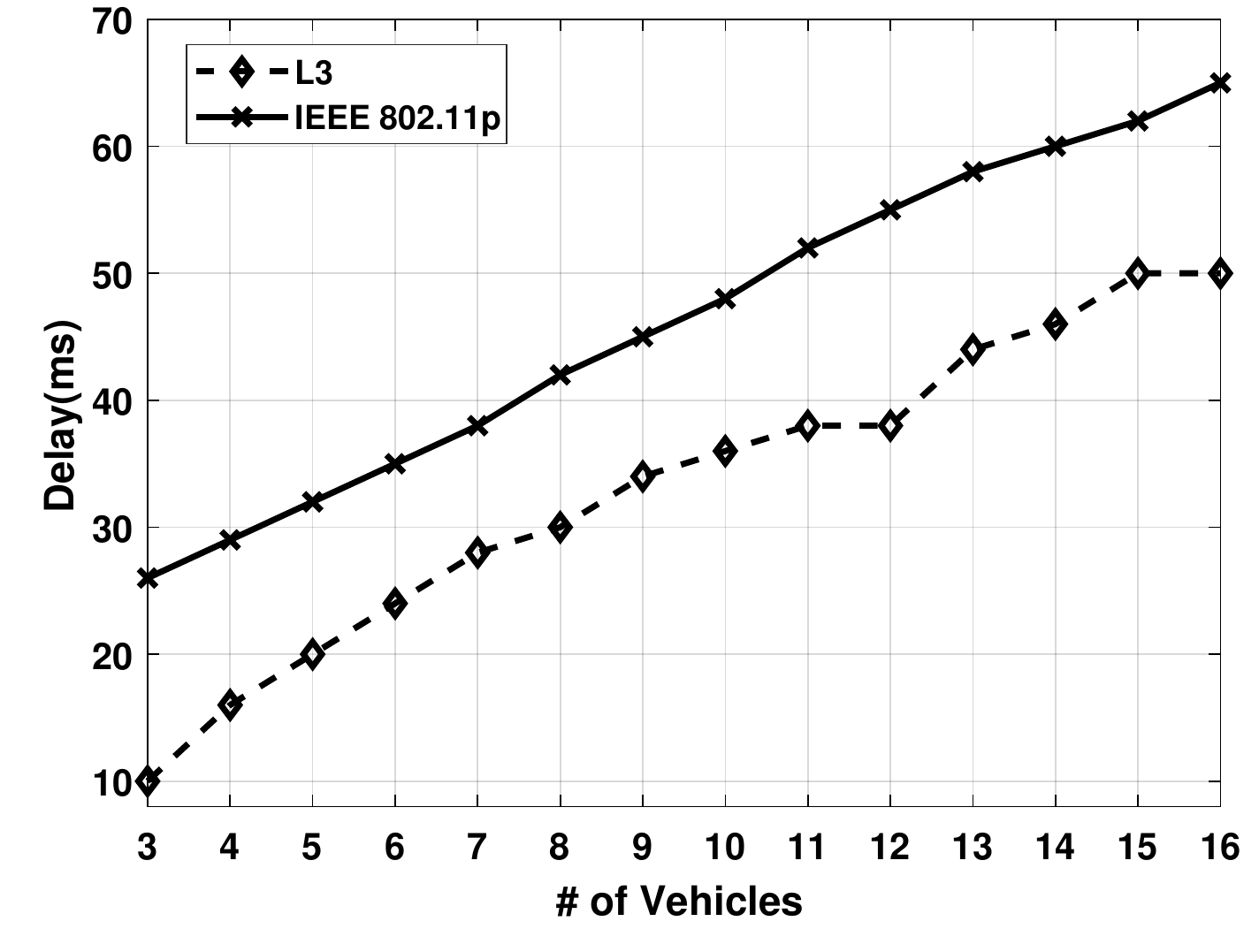}
\caption{Network delay with different numbers of vehicles.}
\label{figure:latency}
\end{figure}

Networking latency of the L3 protocol highly depends on the setting of the time slot, i.e., longer the time slot, larger the networking delay.
To achieve a low-latency protocol, it is critical to set the time slot as small as possible.
To identify the best setting of time slot, we need to understand how long it takes to transmit, receive and process 100 bytes of data in a vehicular network.
We adopt NS-3 simulator~\cite{ns3} to find the minimal required time to finish each round of data transmission between vehicles.
To obtain an accurate measurement of the time, we simulate two vehicles (100m away from each other) communicating to each other in NS-3.
In the simulation, one vehicle transmits a 100-byte message to another vehicle, using the IEEE 802.11p protocol with CSMA/CA disabled.
In this case, the time needed to transmit and receive a 100-byte message is similar to that obtained from capture effect.
Here, the time is what is needed for the receiving vehicle to successfully receive the message from the transmitting vehicle.
In our simulation, it requires less than 2ms to share a 100-byte message between two vehicles.
When the vehicles are closer to each other, the time will be a bit smaller, due to a shorter propagation delay that is neglected in this paper.
As such, we configure the time slot to be 2ms in our simulations.
Fig.~\ref{figure:latency} shows the actual delay of the data sharing process, with different numbers of vehicles in the simulations.
In the figure, there is a notable improvement on latency in L3 over the IEEE 802.11p.
This is because the IEEE 802.11p protocol requires vehicles to compete to access the wireless channels, which may cause a significant networking delay. 

\subsection{Scalability}
\begin{figure}
\centering
\includegraphics[width=0.45\textwidth]{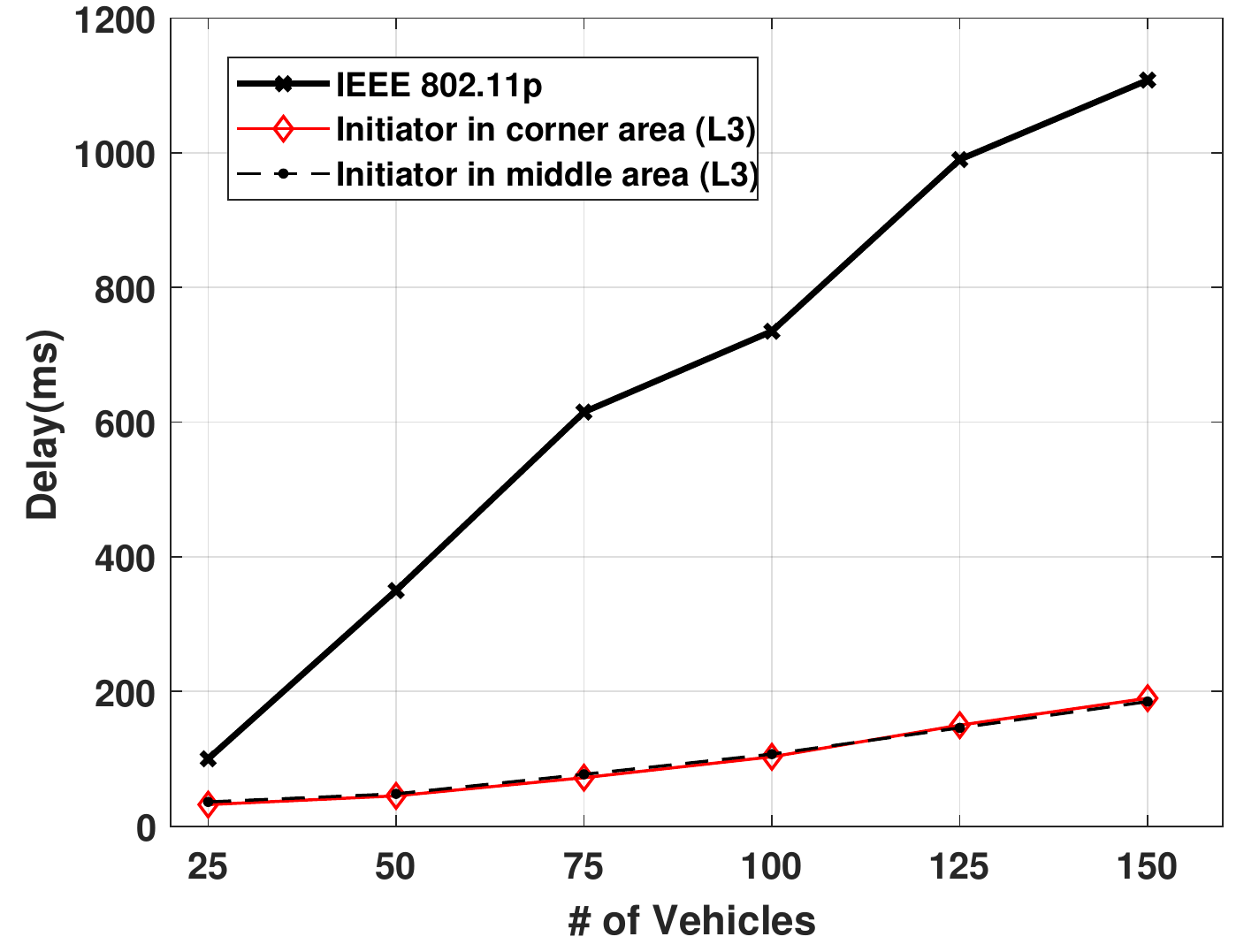}
\caption{Network delay in a large scale vehicular network.}
\vspace{-20pt}
\label{figure:simu}
\end{figure}

With more vehicles, the data sharing among vehicles may take a longer time to complete.
In this section, we conduct experiments to evaluate L3's scalability, i.e., understanding how L3 performs when the number of vehicles increases in the network.
As seen in Fig. \ref{figure:simu}, we witness that the network delay of L3 is increases slightly as the number of participating vehicles increases. 
On the other hand, the latency of IEEE 802.11p tends to perform poorly when there are large number of vehicles transmitting packets simultaneously. 
As vehicles benefiting from the shared data and not suffering from the consequences of large latency, L3 is proved to be effective for up to as many as 225 vehicles in a sensing zone.
With the current traffic infrastructure, there is usually less than 225 vehicle within any reasonable intersection in any city. 
In some extremely crowded areas, the number of vehicles could be larger than 255, which may cause a longer network latency.
To address this issue, we could reduce the size of each sensing zone to include no more than 225 vehicles.
The parameter setting of L3 protocol is not static and needs to be changed based on real-world applications.

\section{Related Work}
Object detection failures and visual obstructions are both core difficulties that all autonomous vehicle must face. 
Techniques such as cooperative perception (COOPER)~\cite{cooper} and others address this problem from a fundamental level through fusion.
While detection results are improved, the wireless bandwidth available for V2X communications is too limited to support huge amount of data transmission among vehicles.

Currently-known fusion methods for connected and autonomous vehicles are categorized into three types: low-level, middle-level and high-level fusions. 
Low-level fusion is also called raw data fusion in which the original sensing data produced by vehicles are transmitted and shared among vehicles~\cite{lowfuse,cooper}.
While middle-level fusion methods make use of the extracted features from raw data to conduct fusion~\cite{midfuse}, high-level fusion mainly combines the sensing results processed by individual vehicles~\cite{highfuse}.
Other approaches like~\cite{pointfusion} and~\cite{f-pointnet}, marry the different sensors from the same vehicle to improve their object detection accuracy.

With the current works detailing the ground work, we know that communication in between vehicles plays an important limiting role based on the type of fusion methods being utilized. 
Taking COOPER~\cite{cooper} for example, while this method improves detection by merging point cloud data, it is limited by the narrow bandwidth available in vehicular networks. 
Not only does using higher quality sensors increase the amount of data that gets generated, using higher quality data also posses the risk that the data being generated will be too big to be transferred efficiently.
Works exploring the sharing data between autonomous vehicles such as~\cite{v1},  discusses the uses of implementing V2X and identifies the requirements for doing so.
The fundamental issue here is that existing vehicular network standards are design to exchange short messages among vehicles, rather than sensing data which could potentially be very large.
%


\section{Conclusions}
We propose the L3 protocol to support low-latency data sharing among autonomous vehicles towards the goal of a better detection of objects around autonomous driving cars.
%
Due to the capture effect, all vehicles transmit their sensing matrices simultaneously and thus a low-latency data sharing among vehicles is achieved.
%
Although the design of L3 protocol is verified and evaluated in simulations, it is worth noting that the implementation of L3 on real-world hardware is still a challenging problem.
In the future, we will explore the possibility of integrating L3 into existing DSRC devices and demonstrate how L3 works in real-world experiments.
\ifCLASSOPTIONcompsoc
  \section*{Acknowledgments}
\else
  \vspace{-7pt}
  \section*{Acknowledgment}
\fi
The work is partially supported by National Science Foundation (NSF) grants NSF CNS-1761641 and CNS 1852134.

%
\bibliographystyle{abbrv}


\bibliography{sigproc}  
%
%

\end{document}